\documentclass[%
 reprint,
 amsmath,amssymb,
 aps,
prd
]{revtex4-1}

\usepackage{lmodern}
\usepackage{graphicx}
\usepackage{dcolumn}
\usepackage{bm}
\usepackage{hyperref}
\hypersetup{linktocpage,colorlinks,citecolor={blue},pdfdisplaydoctitle=true,pdfpagemode=UseOutlines,bookmarksnumbered=true}
\usepackage{mathrsfs,dsfont}
\usepackage{color}
\usepackage{ upgreek }
\usepackage{slashed}

\newcommand{\bra}[1]{\langle #1 |} 
\newcommand{\ket}[1]{| #1 \rangle } 

\definecolor{cbl}{rgb}{0,0,1}

\definecolor{crd}{rgb}{1,0,0}
 
\newcommand{\upd}{\mathrm{d}}
\newcommand{\tr}{\mathrm{tr}}
\newcommand{\xb}{\mathbf{x}}
\newcommand{\yb}{\mathbf{y}}
\newcommand{\pb}{\mathbf{p}}

\newcommand{\im}{\Im \textrm{m}}

\newcommand\densi{{\hat{\upvartheta}}}

\newcommand{\ie}[0]{\textit{i.e.} }
\newcommand{\eg}[0]{\textit{e.g.} }

\begin{document}
\title{Interacting quantum field theories as relativistic statistical field theories of local beables
}

\author{Antoine Tilloy}
\email{antoine.tilloy@mpq.mpg.de}
\affiliation{Max-Planck-Institut f\"ur Quantenoptik, Hans-Kopfermann-Stra{\ss}e 1, 85748 Garching, Germany}
\date{\today}
\begin{abstract}
We propose a reformulation of quantum field theory (QFT) as a relativistic statistical field theory. This rewriting embeds a collapse model within an interacting QFT and thus provides a possible solution to the measurement problem. Additionally, it relaxes structural constraints on standard QFTs and hence might open the way to future mathematically rigorous constructions as well as new numerical methods. Finally, because it shows that collapse models can be hidden within QFTs, this article calls for a reconsideration of the dynamical program, as a possible underpinning rather than as a modification of quantum theory. 
\end{abstract}
\maketitle

\section{Introduction}
In its orthodox acceptation, quantum mechanics is not the dynamical theory of a tangible world. It provides accurate predictions about the results of measurements, but leaves the reality of the microscopic substrate supporting their emergence unspecified. The situation is no different, apart from additional technical subtleties, in the relativistic regime. Quantum field theory (QFT) is indeed no more about fields than non-relativistic quantum mechanics is about particles. At best these entities are intermediary mathematical objects entering in the computation of probabilities. They cannot, even in principle, be approximate representations of an underlying physical reality. More precisely, a QFT (even suitably regularized) does not a priori yield a probability measure on fields living in space-time \footnote{The closest known way to obtain such a representation is through a Wick rotation from Minkowski to Euclidean space-time. But, if the latter has proved to be a useful tool at the axiomatic level, it is of no help to understand QFT in realistic terms.}, even if this is a picture one might find intuitively appealing. 

This does not mean that the very existence of tangible matter is made impossible, but rather that the formalism remains agnostic about its specifics. It seems that some physicists would want more and it is uncontroversial that it would sometimes be helpful to have more (if only to unequivocally solve the measurement problem \cite{maudlin1995,adler2003}). One would likely feel better with local beables \cite{bell1976} (or a primitive ontology \cite{allori2014,allori2015}), \ie with something in the world, some physical ``stuff'', that the theory is about and that can ultimately be used to derive the statistics of measurement results. In the non-relativistic limit, Bohmian mechanics \cite{bohm1952I,bohm1952II,durr2009,goldstein2016} has provided a viable proposition for such an underlying physical theory of the quantum cookbook \cite{durr1992,durr2004}. It may not be the only one nor the most appealing to all physicists, but at least it is a working proof of principle. In QFT, finding an underlying description in terms of local beables has proved a more difficult endeavour. Bohmian mechanics can indeed only be made relativistic in a weak sense \cite{durr2013} (respecting the letter and not the spirit of relativity) and its extension to QFT is sublte \cite{durr2004qft,deckert2016}. At present, there does not seem to exist a fully relativistic theory of local beables that reproduces the statistics of QFT (even setting aside the technicalities of renormalization), although some ground work has been done \cite{kent2013}. The first objective of this article is to propose a solution to this problem and provide a reformulation (or interpretation) of QFT as a Lorentz invariant statistical field theory (where the word ``field'' is understood in its standard ``classical'' sense). For that matter, we shall get insights from another approach to the foundations of quantum mechanics: the dynamical reduction program.

The idea of dynamical reduction models \footnote{Other terminologies include ``spontaneous'' or ``objective'' combined with ``localization'' or ``collapse''. } is to slightly modify the linear state equation of quantum mechanics to get definite measurement outcomes in the macroscopic realm, while only marginally modifying microscopic dynamics. Pioneered by Ghirardi, Rimini, and Weber \cite{ghirardi1986}, Di\'osi \cite{diosi1989}, Pearle \cite{pearle1989,ghirardi1990}, and Gisin \cite{gisin1984} (among others), the program has blossomed to give a variety of non-relativistic models that modify the predictions of the Standard Model in a more or less gentle way. The models can naturally be endowed with a clear primitive ontology, made of fields \cite{bedingham2014}, particles \cite{bedingham2011hv,tumulka2011} or flashes \cite{esfeld2014}. Some instantiations of the program, such as the Continuous Spontaneous Localization (CSL) model \cite{pearle1989,ghirardi1990} or the Di\'osi-Penrose (DP) model \cite{diosi1989,ghirardi1990dp,penrose1996} are currently being put under experimental scrutiny. Like Bohmian theories, these models have been difficult to extend to relativistic settings despite recent advances by Tumulka  \cite{tumulka2006}, Bedingham \cite{bedingham2011} and Pearle \cite{pearle2015}. For subtle reasons we shall discuss later, these latter proposals, albeit crucially insightful for the present inquiry, are difficult to handle and not yet entirely satisfactory. The second objective of this article is thus to construct a theory that can be seen as a relativistic dynamical reduction model while keeping a transparent operational content.

The two aforementioned objectives --redefining a QFT in terms of a relativistic statistical field theory and constructing a fully relativistic dynamical reduction model-- shall be two sides of the same coin. Indeed, our dynamical reduction model will have an important characteristic distinguishing it from its predecessors: its empirical content will be the same as that of an orthodox interacting QFT, hence providing a potential \emph{interpretation} rather than a modification of the Standard Model. This fact may be seen as a natural accomplishment of the dynamical program, yet in some sense also as a call for its reconsideration. Surely, if a dynamical reduction model that is arguably more symmetric and natural than its predecessors can be fully hidden within the Standard Model, it suggests that the ``collapse'' manifestations currently probed in experiments are but artifacts of retrospectively unnatural choices of non-relativistic models.

We should finally warn that the purpose of the present article should not be seen as only foundational or metaphysical. The instrumentalist reader, who may still question the legitimacy of a quest for ontology on positivistic grounds, might nonetheless be interested in its potential mathematical byproducts. As we shall see, because it relaxes some natural constraints on the regularity of QFTs, our proposal might indeed be of help for future mathematically rigorous constructions. Further, the stochastic unraveling tools introduced may be of help for Monte-Carlo simulations of orthodox QFTs.

The article is structured as follows. We first introduce non-relativistic collapse models in section \ref{sec:csl} to gather the main ideas and insights needed for the extension to QFT. The core of our new definition of QFT is provided in section \ref{sec:construction}. We show that the theory allows to understand the localization of macroscopic objects providing a possible natural solution to the measurement problem in section \ref{sec:world}. Finally, we discuss in section \ref{sec:discussion} the implications for QFT and the dynamical reduction program, as well as the limits and the relation to previous work, of our approach.

\section{Non-relativistic dynamical reduction}\label{sec:csl}
There exists numerous clear introductions to dynamical reduction models in the literature (see \cite{bassi2003,bassi2013} for reviews and \cite{ghirardi2016} for an up to date conceptual introduction), but it is worthwhile to remind the reader of their basic features here, especially as we shall put the emphasis on aspects sometimes considered accessory. 

\subsection{Basics}\label{sec:basics}
Although the earliest and perhaps simplest collapse model of Ghirardi, Rimini, and Weber (GRW) \cite{ghirardi1986} is discrete, it makes more sense for the upcoming analysis to start by considering directly its continuous avatar, the mass proportional CSL model \cite{pearle1989,ghirardi1990,bassi2013}. Its so called ``linear'' equation is defined in the following way:
\begin{equation}\label{eq:csllinear}
\begin{split}
\frac{\upd }{\upd t} \ket{\psi_w(t)} = \bigg\{- i \hat{H}_0 + \sqrt{\gamma}& \int_{\mathds{R}^3}\!\! \upd \xb \; \, \hat{M}_\sigma(\xb)  \,w_t(\xb) \\
&- \frac{\sqrt{\gamma}}{2} \hat{M}_\sigma^2(\xb) \bigg\} \ket{\psi_w(t)},
\end{split}
\end{equation}
where $\hat{H}_0$ is the usual non-relativistic Hamiltonian (\eg Schr\"odinger or Pauli-Dirac), $\hat{M}_\sigma(\xb)$ is the $\sigma$-smeared mass density operator, which reads for a system of $N$ distinguishable non-relativistic particles of mass $(m_1,\cdots,m_N)$ and coordinates $(\yb_1,\cdots,\yb_N)$: 
\begin{equation}
\hat{M}_\sigma(\xb)= \frac{1}{\sigma^3 (2\pi)^{\frac{3}{2}}} \sum_{k=1}^N\,\int_{\mathds{R}^3}\!\! \upd \yb_k \, m_k\,  e^{-\frac{|\xb-\yb_k|^2}{2\sigma^2}}\ket{\yb_k}\bra{\yb_k}, 
\end{equation}
and $w_t(\xb)$ is a white noise in space and time $\mathds{E}[w_t(\xb) w_s(\yb)]=\delta^3(\xb-\yb) \delta(t-s)$. The collapse parameter $\gamma$ (sometimes defined divided by $m_0^2$ where $m_0$ is the electron mass) is taken to be ``small'' in the sense that a single isolated particle barely feels the collapse terms. Before discussing how an equation of the kind \eqref{eq:csllinear} might give a plausible solution to the measurement problem, let us first see that its noise-averaged evolution is well behaved. Defining $\hat{\rho}(t)=\mathds{E}_w\big[\ket{\psi_w(t)}\bra{\psi_w(t)}\big]$, we get that
\begin{equation}\label{eq:markovianunraveling}
\frac{\upd }{\upd t} \hat{\rho}(t) = -i\, [\hat{H}_0,\hat{\rho}(t)] - \frac{\gamma}{2}\int_{\mathds{R}^3}\!\! \upd \xb  \left[\hat{M}_\sigma(\xb),[\hat{M}_\sigma(\xb),\hat{\rho}(t)]\right],
\end{equation}
that is, $\rho_t$ has a legitimate open quantum evolution of the Lindblad form. This is central because, at the emergent operational level, equation \eqref{eq:markovianunraveling} will contain all the empirical content of the model. The linearity then guarantees that the theory is free of major inconsistencies like faster than light signalling or break down of the Born rule.
Such an effective evolution \eqref{eq:markovianunraveling} can be obtained by coupling a quantum system to a bosonic bath in the Markov approximation \cite{breuer2002}. This means that the CSL model can be seen as a so called stochastic \emph{unraveling} of a Markovian open system evolution (conversely, all reasonable Markovian collapse models are unravelings of orthodox open evolutions \cite{bassi2013unique}). This fact, which may look trivial at first sight, will subsequently give us important hints in the QFT context.

Now, going back to the collapse equation \eqref{eq:csllinear}, we notice that it generates a non-Hermitian evolution that does not preserve the norm of the state vector, making its interpretation difficult. However, we can simultaneously normalize the state and change its probability measure $\mu_0(w)$, an operation known as Girsanov's transformation \footnote{In the context of collapse models, this operation is sometimes called ``cooking'' of the measure, or passing from the \emph{ostensible} to the \emph{physical} distribution \protect\cite{gambetta2002,gambetta2003}.}:
\begin{align}
    \ket{\widetilde{\psi}_w(t)} &= \frac{\ket{\psi_w(t)}}{\sqrt{\langle \psi_w(t)| \psi_w(t)\rangle}}\\
    \upd \mu_t(w) &=\langle \psi_w(t)| \psi_w(t)\rangle \cdot \upd \mu_{0} (w)
\end{align}
to find an equivalent non-linear evolution that is norm-preserving. Under the new measure $\upd\mu_t$ one can show (see \eg \cite{jacobs2006}) that $\forall \, s\leq t$:
\begin{equation}\label{eq:doob}
w_s(x)=2\sqrt{\gamma}\, \bra{\widetilde{\psi}_w(s)} \hat{M}_\sigma(\xb) \ket{\widetilde{\psi}_w(s)} + b_s(\xb),
\end{equation}
where $b_s(x)$ is a white noise process.
This expression shows that $w_t(\xb)$, a random \emph{classical} field, manifestly carries noisy information about the state (it is sometimes called the ``signal'' in a continuous measurement context \cite{wiseman2009}). Notice importantly that the previous equality \eqref{eq:doob} holds for all $s\leq t$, thus $\upd \mu_t$ can be redefined for $t\rightarrow + \infty$ without changing the marginal probability of the field for times before $t$. This nice feature will unfortunately be lost in more general settings. After redefinition of the measure, one can show that $\ket{\widetilde{\psi}_w(t)}$ verifies:
\begin{align}
\frac{\upd }{\upd t} \ket{\widetilde{\psi}_w(t)} =& \bigg\{\!\! - i \hat{H}_0 + \sqrt{\gamma}\! \int_{\mathds{R}^3}\!\!\!  \upd \xb \! \left[\hat{M}_\sigma(\xb)-\langle \hat{M}_\sigma(\xb)\rangle\right] \! b_t(\xb) \nonumber \\
&- \frac{\sqrt{\gamma}}{2} \left[\hat{M}_\sigma(\xb)-\langle \hat{M}_\sigma(\xb)\rangle\right]^2  \bigg\} \ket{\widetilde{\psi}_w(t)}\label{eq:normalized}
\end{align}
with the compact notation $\langle \hat{M}_\sigma(\xb) \rangle = \bra{\widetilde{\psi}_w(t)} \hat{M}_\sigma(\xb) \ket{\widetilde{\psi}_w(t)}$ and the multiplicative noise term is understood in the It\^o convention. Such an equation, which is the standard form of the CSL model, could have been derived directly through other means, like continuous measurement theory \cite{jacobs2006,wiseman2009}. However, such a luxury will not be available in the QFT context and it is thus worthwhile to understand the main steps of the latter pedestrian computation.

\subsection{Collapse, Amplification, primitive ontology}\label{sec:collapseamplification}
There are three steps to check to see that a collapse model like CSL indeed solves the measurement problem: (i) show that it weakly collapses quantum states in an approximate position basis \footnote{Another possibility that we shall mention later is that the choice of the position basis emerges for the macroscopic center of mass through the amplification mechanism itself.} with the correct probability, (ii) show that this effects is dramatically amplified for many-body systems, and (iii) show that one can define a primitive ontology that allows to understand the orthodox formalism as emergent from its dynamics. As these points are discussed at length in the literature \cite{bassi2007,bassi2013}, we shall explain only briefly why they are verified for the CSL model and insist specifically on the third item.

Neglecting the proper Hamiltonian $\hat{H}_0$ and the smearing $\sigma$, one sees that the fixed points of the evolution \eqref{eq:normalized} are eigenvectors of the mass density operator. The evolution thus progressively drives the system towards states that are well localized in space. Better, this localization is compatible with the Born rule. That is, the probability to reach a given eigenstate when $t\rightarrow +\infty$ with the pure collapse evolution is given by the standard Born rule applied to the initial condition. This is seen by noticing the martingale property of the pure collapse evolution:
\begin{equation}
\forall \, t \geq 0,\, \mathds{E}[\bra{\psi(t)} \hat{M}_\sigma(\xb) \ket{\psi(t)}] = \bra{\psi(0)} \hat{M}_\sigma(\xb) \ket{\psi(0)},
\end{equation}
which is a straightforward consequence of \eqref{eq:markovianunraveling}. Another way to understand this progressive collapse according to the Born rule is that \eqref{eq:normalized} is formally equivalent to the continuous non-demolition quantum measurement of the (regularized) mass density. 

The amplification mechanism is no more difficult to understand. We illustrate it on a simplified situation but it should be clear that it is much more general. We consider a system of $N$ bound particles typically constituting the pointer of a measurement apparatus. We assume that the pointer can be in two sharply localized positions $l$ and $r$, separated by a distance $d \gg \sigma$. We consider a macroscopic superposition of the pointer, \ie we consider the total state:
\begin{equation}
\ket{\Psi(0)} \propto \ket{l}\otimes \cdots \otimes \ket{l} + \ket{r} \otimes \cdots \otimes \ket{r} := \ket{L}+\ket{R}.
\end{equation}
The CSL evolution \eqref{eq:csllinear} keeps the state in the basis $\{\ket{L},\ket{R}\}$. The collapse evolution in this latter basis is the same as the evolution for a single particle in the basis $\{\ket{l},\ket{r}\}$ but with a new collapse parameter $\gamma_N = N^2\gamma$. This is the amplification mechanism: because of the factor $N^2$, the collapse effect can be at the same time negligible for single particle dynamics and dominant for large macroscopic superpositions.

It is now important to notice that the two previous points are not in themselves sufficient to unequivocally solve the measurement problem. Indeed, we now have a more intuitive evolution for a quantum state but the latter still needs a physical interpretation. Although it has been argued that the quantum state alone --coding a physical world irreducibly in Hilbert space-- might give a satisfactory picture of reality \cite{ney2013}, introducing local beables (or a primitive ontology) is a simpler way to connect the formalism to the world (and crucial for our subsequent analysis). The standard choice in the literature is to take the mass density $\langle \hat{M}_\sigma(\xb) \rangle$. It is indeed a field in physical space that projects down the localization properties of the wave function in configuration space and thus seems to give an intelligible picture of the world. However, the random field $w_t(\xb)$ is another possibility which, as we shall argue, has a few nicer properties and will be instrumental for our redefinition of QFT. This will be clearer in the upcoming section but we can nonetheless say a few words about it now. First of all, $w_t(\xb)$ is present in the collapse model from the very beginning in the linear collapse equation \eqref{eq:csllinear} and the quantum state is a function of it $w\rightarrow \ket{\widetilde{\psi}_w}$, \ie by considering $w$ only, we do not lose information nor do we introduce a new field. After the change of measure, it is clear from equation \eqref{eq:doob} typically has the same localization properties as $\langle \hat{M}_\sigma(\xb) \rangle$ but for some additional white noise, hence it still gives an intuitive picture of the world. Finally, making $w$ gravitate, which may seem natural for a primitive ontology, it has been showed that one could obtain a consistent semi-classical theory of gravity, at least in the Newtonian regime \cite{tilloy2016}.

In the end, the CSL model can be seen as a dynamical theory for a (classical) random field $w$ where the quantum state is simply a convenient way to store the modification of the field Gaussian probability measure. The configuration of macroscopic objects and especially of measurement pointers can be understood in terms of the spatial configuration of the field. Ultimately, the master equation \eqref{eq:markovianunraveling} is enough to derive predictions at the operator level, yet it is the primitive ontology $w$ that is the fundamental object of the theory and that gives meaning to the whole story. The situation may be understood with an analogy with the kinetic theory of gasses where all the macroscopic predictions can be computed using thermodynamic variables but where it is the particles that are the fundamental real entities, from which the thermodynamic picture only emerges.

\subsection{Extensions and general lessons}\label{sec:extension}
\begin{figure*}
\centering
\includegraphics[width=0.99\textwidth]{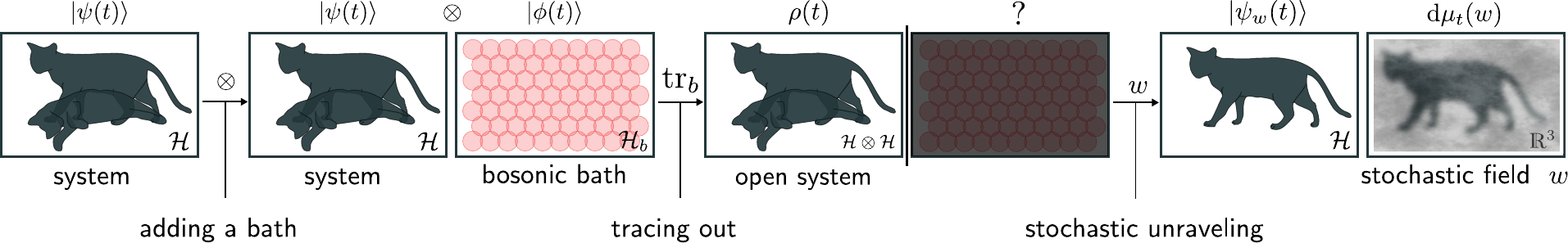}
\caption{General recipe to construct a collapse model. The system is coupled with a bosonic bath which is subsequently traced out and stochastically unraveled. The final product can be considered to be the random field $w$ living in real space. Macroscopic superpositions are collapsed and projected down to $\mathds{R}^3$ but the predictions of the theory are modified. The cat drawing is borrowed from a picture by D. Hatfield \cite{doug}. }
\label{fig:recipe1}
\end{figure*}

Non-relativistic collapse models have been extended in various ways to add non-white noises \cite{bassi2002,adler2007,adler2008}, dissipation effects \cite{smirne2014,smirne2015} or both \cite{ferialdi2012}. These new models may seem even more ad hoc than their historical counterpart, yet they share a unifying feature: they can all be seen as the stochastic unraveling of an open system evolution. Where the CSL model corresponds to the coupling with a Markovian bosonic bath, the non-Markovian and dissipative models correspond to the coupling (linear in the bath field operators) with a \emph{generic} bosonic bath. This insight that a large class of (possibly non-Markovian) stochastic Schr\"odinger equations can be seen as unravelings of this latter kind of open-system evolution has been shown in the clearest way by Di\'osi and Ferialdi \cite{diosi2014} (see Fig. \ref{fig:recipe1} for a recipe of collapse models).

\begin{figure}
\centering
\includegraphics[width=0.99\columnwidth]{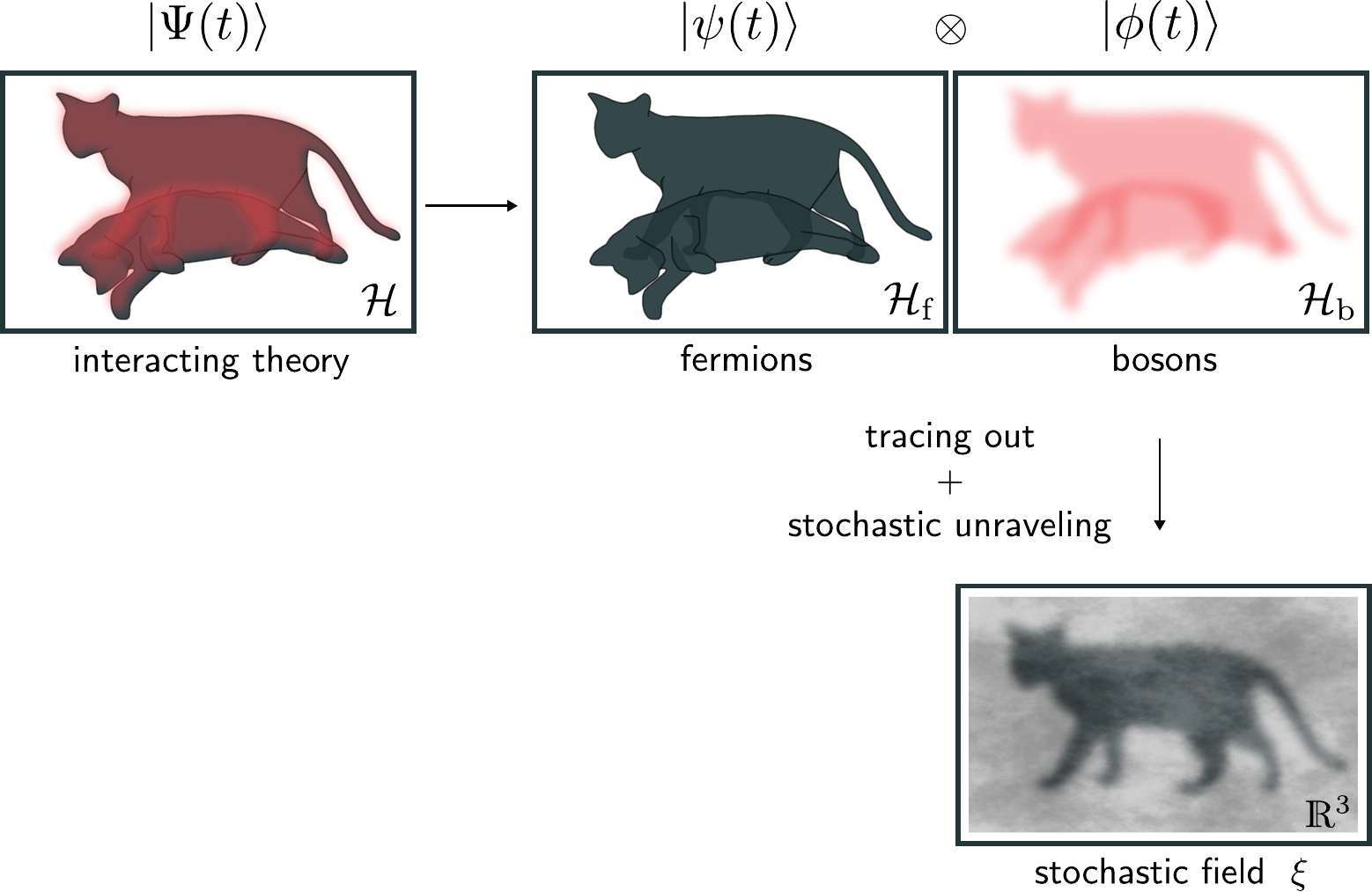}
\caption{Recipe to construct a collapse model, hence a statistical field theory, from an interacting QFT. Separating fermions and bosons, one can take the latter as bath and apply the steps of figure \ref{fig:recipe1}. The final product is the stochastic field $\xi$ and the predictions of the theory are unchanged.}
\label{fig:recipe2}
\end{figure}

The lesson one may draw from this is that a good way to construct dynamical reduction models that are consistent \emph{by design} is as stochastic unravelings of legitimate open-system evolutions. This gives the most important hint for our subsequent extension to QFT. For an interacting QFT of fermions and bosons, there is indeed a good candidate for a legitimate open-system evolution: the reduced evolution of the fermions when the bosons are simply traced out! Before making this precise, notice that we can actually go much further. Instead of adding an ad hoc bath that is subsequently unraveled, it is natural to take a bosonic sector that already exists in the QFT that is considered such as QED in the Standard Model (see Fig. \ref{fig:recipe2}). Although it may be cumbersome in practice, the measurement results on bosons can always be written in terms of macroscopic fermionic observables (and vice versa) \cite{goldstein2005,struyve2007,struyve2010}. One would thus not lose any information in such a rewriting. We can now come to the main idea of this article:
\begin{quotation}
\noindent An interacting QFT of fermions and bosons can be \emph{rewritten} as a (non-Markovian) dynamical reduction model on fermions. With a proper choice of primitive ontology / local beables (discussed in \ref{sec:collapseamplification}), this gives a relativistic statistical theory of classical fields from which the quantum formalism can be shown to emerge.
\end{quotation}
The aim of the next sections is to explicitly do this construction and discuss its numerous theoretical and practical consequences.

\section{Quantum field theories as statistical field theories}\label{sec:construction}
Our objective is to now understand how the previous ideas can be made concrete in the context of QFT. Although there is no fundamental difficulty in carrying the subsequent derivations in a more general context, we will discuss a Yukawa theory of Dirac fermions interacting with scalar bosons for the sake of simplicity. However, the reader should keep in mind that, unless otherwise stated, the results would hold for more realistic fundamental theories like QED.

Further, we will assume that the propagators that will show up are suitably covariantly regularized at a scale $\Lambda$  and thus that the Lagrangian is written with ``bare'' parameters. For reasons we will discuss later, the ability to consider regularized theories as fundamental is an appealing feature of our rewriting (which does not a priori forbid the discussion of renormalization). We should however recognize that we adopt this admittedly naive approach to orthodox QFT also because it is not straightforward to proceed otherwise at that stage. Finally we will use natural units in which $\hbar=c=1$.

The action of our interacting theory is the following:
\begin{equation} 
\mathscr{S} = \int_{\mathds{R}^4} \!\! \upd^4 x \; \, \mathscr{L}_f(\psi, \partial_\mu \psi) + \mathscr{L}_{\mathrm{int}}(\psi,\phi) + \mathscr{L}_b(\phi,\partial_\mu \phi)
\end{equation}
with the Lagrangian densities:
\begin{align}
\mathscr{L}_f(\psi, \partial_\mu \psi)&=\bar{\psi} (i \slashed{\partial} -m_f)\psi \\
\mathscr{L}_{\mathrm{int}}(\psi,\phi) &= g\, \bar{\psi} \, \psi\, \phi\\
\mathscr{L}_b(\phi,\partial_\mu \phi)&= \frac{1}{2}\partial_\mu\phi\, \partial^\mu\phi - \frac{1}{2}m_b^2 \phi^2,
\end{align}
where $m_f$ is the fermionic mass, $m_b$ the bosonic mass and $g$ the coupling constant. We thus assume that the bosonic part of the action does not contain higher powers in the field (like $\phi^4$, that would be needed for perturbative renormalization). Such additional interaction terms can nonetheless be dealt with formally if needed (see appendix \ref{appendix:decoupling}).
We will now redo the steps of section \eqref{sec:csl} both in the operator picture and in the functional integral picture. These two point of views will prove useful to emphasize different structural properties of the theory: the first putting the emphasis on states and dynamical reduction, the second on the local beable field and Lorentz invariance.

\subsection{Operator picture}\label{sec:operator}
We first consider the point of view that an interacting QFT is a theory providing a dynamical law for quantum states and density matrices (or for operators in the Heisenberg picture) at least at a formal level. This is the most convenient approach to emphasize the parallel with dynamical reduction models as the latter are typically understood, albeit perhaps naively as we have previously written, as stochastic modifications of the standard evolution for states. As the functional integral picture will allow for an easier understanding of Lorentz invariance, we delay its discussion to the next subsection and use a unique reference frame with Cartesian coordinates here. We thus simply consider states defined on constant time hypersurfaces. In this framework, and leaving issues of divergences and renormalization aside, a relativistic QFT is no different from a simple non-relativistic quantum theory. Eventually, as is customary in this context, we write all the operators (\ie the objects with a hat) in the interaction picture. 

\subsubsection{Obtaining the linear stochastic unraveling}

Before starting, let us recall our objective: we first want to trace out the bosonic quantum field and then stochastically unravel the corresponding open-system evolution. For that matter the important intermediary object to consider will be the fermionic reduced density matrix $\hat{\rho}_f(t)=\tr_b\left[\hat{\rho}(t)\right]$ obtained by tracing out the bosonic degrees of freedom. Provided the total density matrix is known at time $t_i$, the reduced density matrix is given by:
\begin{equation}\label{eq:defrho}
\hat{\rho}_f(t) = \tr_b \left\{\mathcal{T}\!\exp\left[ -i\!\int_{t_i}^{t} \!\!\upd^4x \, \mathscr{H}^L_{\rm int} (x)-\mathscr{H}^R_{\rm int} (x)\right] \cdot \hat{\rho} (t_i)\right\} 
\end{equation}
where $\mathcal{T}$ is the time ordering operator and we have used the left-right notation for super-operators: $A^L \cdot \hat{\rho} = \hat{A}\hat{\rho}$ and $A^R\cdot \hat{\rho} = \hat{\rho}\hat{A}$. We now want to derive an open-system evolution involving the reduced density matrix $\hat{\rho}_f$ only. For that matter, we need to make the hypothesis that the total initial state is a product of fermionic and bosonic degrees of freedom at some initial time $t_i$ (which we will later be sent to $-\infty$): $\hat{\rho}(t_i) = \hat{\rho}_f(t_i) \otimes \hat{\rho}_b(t_i)$. With this hypothesis, and provided the bosonic state $\hat{\rho}_b(t_i)$ is the vacuum or a thermal state, the bosonic degrees of freedom can be explicitly integrated out (see appendix \ref{appendix:operatorinfluence}) to give:
\begin{equation}\label{eq:influencefunctional}
\hat{\rho}_f(t) = \mathcal{T} \exp \left(i\Phi \left[j^L,j^R\right]\right)\cdot \hat{\rho}_f(t_i)
\end{equation}
where $\Phi [j^L,j^R]$ is the ``operator'' influence phase functional:
\begin{equation}
\begin{split}
i\Phi\left[j^L,j^R\right] =& \int_{t_i}^{t} \!\! \int_{t_i}^{t} \!\!  \upd^4x\, \upd^4y \; D(x,y) \, j^L(x) j^R(y) \\
&- \frac{1}{2} \theta(x^0-y^0) \, D(x,y) \, j^L(x)j^L(y) \\
&- \frac{1}{2}\theta (y^0-x^0) \, D(x,y) \, j^R(x)j^R(y),
\end{split}
\end{equation}
with $D(x,y) = \tr_b \left[\hat{\phi}(x)\hat{\phi}(y) \,\hat{\rho}_b(t_i)\right]$, the two-point correlation function (not time ordered) of the free bosons, and $\hat{\jmath}=g\, \widehat{\bar{\psi} \psi}$.

To obtain an evolution equation for fermionic pure states (assuming the initial fermionic state was pure $\hat{\rho}_f(t_i)=\ket{\Psi(t_i)}\bra{\Psi (t_i)}$), we have to decouple the left and right part of the influence phase functional. For that matter, the crucial fact we shall use is that $ \mathcal{T} \exp \left(i\Phi \left[j^L,j^R\right]\right)$ looks very much like the Fourier transform (or characteristic function) of a complex stochastic Gaussian field (see appendix \ref{appendix:gaussianfields}). Indeed, such a function takes the form of an exponential of a bilinear term involving only the two point correlation functions of the process. Additionally, one can show (see appendix \ref{appendix:gaussianfields}) that there exists a legitimate stochastic field $\xi$ such that $D(x,y) = \tr_b \left[\hat{\phi}(x)\hat{\phi}(y) \,\hat{\rho}_b(t_i)\right] = \mathds{E}_{t_i}[\xi(x)\xi^*(y)]$ where $\mathds{E}_{t_i}[\cdot]=\int \cdot \, \upd \mu_{t_i}(\xi)$ denotes the stochastic average and $\upd \mu_{t_i}$ is the (a priori) measure on $\xi$. This correlation function does not fix completely the field $\xi$ and we need to specify its relation matrix $S(x,y)=\mathds{E}_{t_i}[\xi(x)\xi(y)]$ which will remain a free parameter of the model. With a bit of guess work and following \cite{diosi2014} one gets that the random state:
\begin{widetext}
\begin{equation}\label{eq:psiintegrated}
\ket{\psi_\xi(t)}=\mathcal{T}\exp\left\{-i \int_{t_i}^t \!\! \upd^4x \; \hat{\jmath}(x)\, \xi(x) - \int_{t_i}^{t} \!\! \int_{t_i}^{t} \!\!  \upd^4x\, \upd^4y \,  \theta(x^0-y^0)\, [D-S](x,y)\, \hat{\jmath}(x)\hat{\jmath}(y)\right\} \ket{\psi_\xi(t)}
\end{equation}
is such that the ``unraveling'' condition
\begin{equation}\label{eq:unraveling}
\mathds{E}_{t_i}\Big[\ket{\psi_\xi(t)}\bra{\psi_\xi(t)}\Big] = \rho_f(t)
\end{equation}
 is fulfilled. This latter equality is proved by Gaussian integration or equivalently through the use of Wick's theorem (see appendix \ref{appendix:gaussianfields}). Equation \eqref{eq:psiintegrated} may be written in the perhaps more familiar differential form \cite{diosi1997,diosi2014}:
\begin{equation}\label{eq:psidifferential}
\frac{\upd}{\upd t} \ket{\psi_\xi(t)} = - i \left\{\int_{\mathds{R}^3}\!\! \upd^3\mathbf{x} \; \hat{\jmath}(t,\mathbf{x})\left[\xi(t,\mathbf{x}) + \int_{t_i}^t \!\! \upd^4 y \, [D-S]\left((t,\mathbf{x}),(y^0,\mathbf{y})\right) \frac{\delta}{\delta \xi(y)}\right] \right\} \ket{\psi_\xi(t)}.
\end{equation}
\end{widetext}
Although this latter kind of equation is commonly used in the literature on stochastic Schr\"odinger equations, one should bear in mind that \eqref{eq:psiintegrated} guarantees that $\ket{\psi_\xi}$ can in theory be computed from a \emph{single} realization of $\xi$, something one might naively put in doubt \footnote{Indeed, the functional derivative suggests that it is necessary to know ${|\psi_\xi\rangle}$ for a given $\xi$ but also for an ensemble of neighboring realizations \protect\cite{gambetta2002}. This is fortunately not the case.} knowing only \eqref{eq:psidifferential}. Notice that this equation is the direct non-Markovian analog of the linear collapse equation \eqref{eq:csllinear} and that $\xi$ is the QFT counterpart of the random field $w$ of section \ref{sec:basics}.

\subsubsection{Redefining the field measure}\label{sec:fieldmeasure}
As before, the state is not yet normalized at that stage and we need to redefine it and Girsanov transform the measure to preserve the unraveling condition \eqref{eq:unraveling}:
\begin{align}
    \ket{\widetilde{\psi}_\xi(t)} &= \frac{\ket{\psi_\xi(t)}}{\sqrt{\langle \psi_\xi(t)| \psi_\xi(t)\rangle}}\\
    \upd \mu_t(\xi) &=\langle \psi_\xi(t)| \psi_\xi(t)\rangle \cdot \upd \mu_{t_i} (\xi).\label{eq:cooking}
\end{align}
At this stage, there is an important subtlety compared to the Markovian case. In this non-Markovian context, it is not true anymore that the marginal probability of the field $\xi(s,\xb)$ at some time $s\leq t_1<t_2$ is independent of the choice of the measure $\upd \mu_{t_1}$ or $\upd \mu_{t_2}$. That is, the state $\ket{\widetilde{\psi}_\xi(s)}$, where $\xi$ has measure $\upd \mu_t(\xi)$ for $t>s$, does not verify the unraveling condition \eqref{eq:unraveling}.
The standard way to circumvent this difficulty, used to define non-linear non-Markovian stochastic Schr\"odinger equations \cite{diosi1998,strunz1999,gambetta2002} (and thus standard non-Markovian collapse models as well) is to redefine the field variable dynamically $\xi \rightarrow \xi^{[t]}$ in such a way that:
\begin{equation}\label{eq:changevariable}
\left|\text{det} \left[\frac{\delta \xi^{[t]}}{\delta \xi}\right]\right|\, \upd \mu_t(\xi^{[t]}(\xi)) = \upd \mu_{t_i} (\xi).
\end{equation}
That is, $\xi$ is first drawn from the a priori measure $\upd \mu_{t_i}(\xi)$ and, at each time $t\geq t_i$, a new complete field configuration $\xi^{[t]}$ is constructed in such a way that it has the correct measure. Indeed one then has:
\begin{align}
\int_\xi \cdot\; \upd \mu_{t_i} (\xi)&=\int_\xi \cdot\, \left|\text{det} \left[\frac{\delta \xi^{[t]}}{\delta \xi}\right]\right|\, \upd \mu_t(\xi^{[t]}(\xi))\\
&=\int_{\xi^{[t]}} \!\!\!\!\cdot\; \upd \mu_{t}(\xi^{[t]}).
\end{align}
The random state $t\rightarrow \ket{\widetilde{\psi}_{\xi^{[t]}}(t)}$ consequently verifies the unraveling condition \eqref{eq:unraveling} at \emph{all times} and is taken as the output of the procedure. The random state $\ket{\widetilde{\psi}_{\xi^{[t]}}(t)}$ has a priori no nice symmetry property, and depends upon the foliation of space-time that is chosen. However, what matters for us is the associated field of local beables. The ``good'' choice is to take:
\begin{equation}
    \widetilde{\xi}: (t,\xb) \longmapsto \xi^{[t_f]}(t,\xb),
\end{equation}
where $t_f$ is the final time of the evolution (later sent to $+\infty$). The field $\widetilde{\xi}$, just like the field $w$ of the CSL model, carries information about the state and is not anymore a Gaussian field of zero mean. Indeed, for the special choice $S=0$, it is possible to show using \eqref{eq:changevariable} (see Appendix \ref{appendix:girsanov} ) that:
\begin{equation}\label{eq:fieldfield}
\begin{split}
\widetilde{\xi}(t,\xb)= \xi(t,\xb)+ &i \int_{t_i}^{t_f} \!\!\upd s \! \int_{\mathds{R}^3}\!\!\upd^3 \yb \, D((t,\xb),(s,\yb)) \\
&\times \bra{\widetilde{\psi}_{\xi^{[s]}}(s)} \hat{\jmath}(s,\yb) \ket{\widetilde{\psi}_{\xi^{[s]}}(s)},
\end{split}
\end{equation}
which is the exact analogue of equation \eqref{eq:doob} obtained for the CSL model.
Consequently, as before, if and when the state $\ket{\widetilde{\psi}_{\xi^{[t]}}(t)}$ does collapse, the beable field will contain this information. 

At that point, it is important to emphasize that the standard way to define the noise field in the literature is to take $\check{\xi}(t,\xb)\mapsto \xi^{[t]}(t,\xb)$ instead of $\widetilde{\xi}: (t,\xb)\mapsto \xi^{[t_f]}(t,\xb)$. For such a choice, $\check{\xi}$ is obtained from $\xi$ via equation \eqref{eq:fieldfield} but for the substitution $t_f\rightarrow t$ in the bound of the first integral. On the example of a non-relativistic bosonic bath, Gambetta and Wiseman have shown \cite{gambetta2003} that the analog of the field $\check{\xi}$ could be interpreted as a hidden variable associated to the bath, where the (modified) equation \eqref{eq:fieldfield} is interpreted as a guiding equation for bath hidden variables in a modal interpretation of quantum mechanics. Such an interpretation is lost in our case for $\widetilde{\xi}$: as a function of the random state, $\widetilde{\xi}$ looks like it is guided from the past \emph{and} from the future. At that stage, this may look like an unnecessary complication. Actually as we shall see, although the decomposition of $\widetilde{\xi}$ in two terms given by \eqref{eq:fieldfield} is manifestly frame dependent, the probability measure of $\widetilde{\xi}$ taken as a whole is not. This is a crucial advantage of our choice of beable field over the standard noise field.

Eventually, the collapse model we obtain is contained in the stochastic evolution $\ket{\widetilde{\psi}_{\xi^{[t]}}(t)}$ which is frame dependent and in the beable field $\widetilde{\xi}$ which, again as we shall see, has a relativistic probability distribution. The stochastic states should be understood as a tool one can use to understand the theory or make computations. The field $\widetilde{\xi}$, on the other hand, is what we take as the final objective product. This suggests to now \emph{read} the interacting QFT we started from as a statistical field theory of a complex scalar field $\widetilde{\xi}$ with measure $\upd \mu_{t_f} (\widetilde{\xi})$, and with $t_i$ pushed to $-\infty$ and $t_f\rightarrow +\infty$ to get the field defined on all space-time. This is the choice we shall now make. This operator-state picture makes it perfectly clear that we strictly follow the non-relativistic steps of \ref{sec:csl} with the guidelines suggested in \ref{sec:extension}. Our construction also makes it also rather intuitive that the collapse (according to the Born rule) and amplification phenomena discussed in \ref{sec:collapseamplification} will follow as well in this more general context. Before going beyond this heuristic understanding, we have to discuss another perspective that puts the field $\xi$ at the center of the game to show, as promised, that its probability distribution is given in a rather natural way by a relativistic statistical field theory.

\subsection{Functional integral picture}\label{sec:functional}
As advertised, we now forgo the state (or operator) description to focus on the field $\tilde{\xi}$ defined before. For that matter, we use the functional point of view of QFT in the so called ``in-in'' formalism. We also use boundary conditions pushed at $\infty$ right from the start to simplify the presentation. In this framework, the fermionic density matrix on some asymptotic hyper-surface reads:
\begin{equation}
\begin{split} 
\rho_f(\Psi_{+\infty},\Psi'_{+\infty})= \int &\mathcal{D}[\Psi] \mathcal{D}[\Psi']  \mathcal{D}[\phi]  \mathcal{D}[\phi']  \, 
\rho_b(\phi_{-\infty},\phi'_{-\infty})  \\ &  \rho_f(\psi_{-\infty},\psi'_{-\infty})\exp\left( i\mathscr{S} - i\mathscr{S}'\right), 
\end{split}
\end{equation}
 where we have used the simplified notation $\Psi=(\psi,\bar{\psi})$. The functional integral goes through all fields $\Psi$ and $\Psi'$ asymptotic to $\Psi_{+\infty}$ and $\Psi'_{+\infty}$, and all $\phi$ and $\phi'$ such that $\phi_{+\infty}=\phi'_{+\infty}$ (which corresponds to taking the trace).
As before we want to integrate out the bosonic degrees of freedom, that is to compute the influence functional:
\begin{equation}
\begin{split}
\mathcal{F}[j,j']:=\int& \!  \mathcal{D}[\phi]\mathcal{D}[\phi']\,\delta(\phi_{+\infty},\phi'_{+\infty}) \rho_b(\phi_{-\infty},\phi'_{-\infty})\\
&\exp\left[ i\left(\mathscr{S}_b+\mathscr{S}_{\rm int}\right) - i\left(\mathscr{S}_b'+\mathscr{S}_{\rm int}'\right)\right].
\end{split}
\end{equation}
One can show that the influence functional can be expressed exactly in the same way as before \eqref{eq:influencefunctional} as $\mathcal{F}[j,j']=\exp\left(i\Phi[j,j']\right)$ with a quadratic influence phase functional:
\begin{equation}\label{eq:functionalphase}
\begin{split}
i\Phi\left[j,j'\right] =& \iint \! \upd^4x\, \upd^4y \; D(x,y) \, j(x) j'(y) \\
&- \frac{1}{2} \theta(x^0-y^0) \, D(x,y) \, j(x)j(y) \\
&- \frac{1}{2}\theta (y^0-x^0) \, D(x,y) \, j'(x)j'(y).
\end{split}
\end{equation}
This form is again valid if the initial bosonic state is the vacuum or a thermal state asymptotically in the past. As our objective is to build a potentially fundamental theory, we will focus on Lorentz invariant propagators and thus momentarily neglect the possibility of thermal states.

\begin{widetext}
We can decouple the $j j'$ term as the average over a complex (but \emph{classical}) Gaussian stochastic field $\xi$ as in \eqref{eq:psiintegrated}:
\begin{equation}
\begin{split}
\mathcal{F}[j,j' ] =& \mathds{E}_{-\infty}\left[ \mathcal{Q}[j] \mathcal{Q}^*[j']\, \exp \left\{-i \int \upd^4 x\;  j(x)\,\xi(x) - j'(x) \xi^*(x) \right\}\right],  
\end{split}
\end{equation}
with
\begin{align}
    \mathcal{Q} \, [j\, ]&=\exp\left[\frac{-1}{2} \iint \!  \upd^4x\, \upd^4y \,   \theta(x^0-y^0) \, [D-S](x,y) \, j\, (x)\, j\, (y) \right].
\end{align}
At that point, we require that $S$ be also a Lorentz invariant kernel so that $\mathcal{Q}[j]$ stays manifestly frame independent. To make things more explicit in this functional integral context, we can write the stochastic Gaussian measure as a functional integration measure:
\begin{equation}
\upd \mu_{-\infty}(\xi) \propto \mathcal{D}[\xi] \mathcal{D}[\xi^*]\, e^{- \mathscr{S}_{\mathrm{sto}} [\xi,\xi^*]},
\end{equation}
where $\mathscr{S}_{\mathrm{sto}} [\xi,\xi^*]$ is a positive semi-definite quadratic action functional (see appendix \ref{appendix:gaussianfields}),
\begin{equation}
\mathscr{S}_{\mathrm{sto}} [\xi,\xi^*]=\frac{1}{2} \iint\! \upd^4x \, \upd^4 y \, \left(\xi^*(x), \xi(x)\right) \left(\begin{array}{cc}
       D & S  \\
       S^* &D^* 
    \end{array}\right)^{-1}\!\!\!\!\!(x,y) \left(\begin{array}{c}
          \xi(y) \\
         \xi^*(y) 
    \end{array}\right) 
\end{equation}
where the inverse is taken in the operator sense. This stochastic action functional is also a manifestly Lorentz invariant object. The redefinition of the field measure can now be done as in equation \eqref{eq:cooking} and one obtains:
\begin{equation}\label{eq:fieldcookedmeasure}
\begin{split}
\upd\mu_{+\infty}(\widetilde{\xi})=&\int \! \mathcal{D}[\Psi]  \mathcal{D}[\Psi']\, \rho_f(\Psi_{-\infty},\Psi'_{-\infty})\,\delta(\Psi_{+\infty},\Psi'_{+\infty}) 
\mathcal{Q}^*[j'] \, \mathcal{Q}[j] \exp\left(i \mathscr{S}_f-i\mathscr{S}_f' -i\int j \widetilde{\xi}-j'\widetilde{\xi}^*\right) \upd \mu_{-\infty} (\widetilde{\xi}).
\end{split}
\end{equation}
This concludes the definition of our reformulation of a quantum field theory as a statistical field theory for $\widetilde{\xi}$, where \eqref{eq:fieldcookedmeasure} can now be taken as the starting point \footnote{In its definition \eqref{eq:fieldcookedmeasure} it incidentally takes a form that resembles, at least cosmetically, the ``real'' path integral idea of Kent \protect\cite{kent2013}.}. It is a field theory in the standard sense of the word as $\upd \mu_{+\infty}$ is a genuine probability measure (at least at a non rigorous level). It is as Lorentz invariant as it could be in the sense that the field probability measure is given by a Lorentz invariant functional, up to boundary condition terms. Naturally, the probability measure of the field $\widetilde{\xi}$ depends explicitly on the initial state: in this framework, asymptotic states of fermions are simply abstract objects that allow to compactly incorporate some knowledge about the past in the field probability measure. The initial condition is encoded in the state $\rho_f(\Psi_{-\infty},\Psi'_{-\infty})$ and the final condition in the identity, or maximally mixed state $\delta(\Psi_{+\infty},\Psi'_{+\infty})$. However, at that point, we may very well decide to allow for a wider class of boundary conditions, if only to make the theory more symmetric. We may thus consider more general measures of the form:
\begin{equation}\label{eq:generalized}
\begin{split}
\upd\mu_{+\infty}(\widetilde{\xi})\propto \!\int \! \mathcal{D}[\Psi]  \mathcal{D}[\Psi']&\, \rho^{\rm in}_f(\Psi_{-\infty},\Psi'_{-\infty})\,\rho^{\rm out}_f(\Psi_{+\infty},\Psi'_{+\infty}) 
\mathcal{Q}^*[j'] \, \mathcal{Q}[j] \exp\left(i \mathscr{S}_f-i\mathscr{S}_f' -i\int j \widetilde{\xi}-j'\widetilde{\xi}^*\right) \upd \mu_{-\infty} (\widetilde{\xi}).
\end{split}
\end{equation}
\end{widetext}
Notice finally that classical Grassmanian sources $J$ and $\bar{J}$ for the fermions can easily be added in the previous formula \eqref{eq:fieldcookedmeasure},\eqref{eq:generalized} and provide yet another way to tilt the probability measure. In the following, we will restrict ourselves to discussing the simplest possibility of \eqref{eq:fieldcookedmeasure} which is the closest in spirit to standard QFT.

\subsection{Computing the statistics of the beable field}
Now that we have a formal probability measure for the field $\widetilde{\xi}$, we can in principle compute all its correlation functions. A first possibility is to do it perturbatively in the coupling constant $g$ (remember that $j$ contains one power of $g$). One just needs to Dyson expand the non-Gaussian part of equation \eqref{eq:fieldcookedmeasure} which will lead to a mild extension of standard diagrammatics. This represents no further conceptual difficulty especially if we use covariantly regularized propagators at a fixed scale $\Lambda$. Actually, as we shall later argue in \ref{sec:reconsideringQFT}, this ability to work with regularized propagators as if they were fundamental is an appealing mathematical consequence of the present formalism. In the end, it is possible to compute all the field correlation functions via a perturbative expansion of the form:
\begin{equation}
\begin{split}
    \mathds{E}\big[\widetilde{\xi}(x_1)\cdots \widetilde{\xi}(x_k)\widetilde{\xi}^*(x_{k+1})&\cdots\widetilde{\xi}^*(x_n)\big] \\
    &= \sum_{\ell=0}^{+\infty} g^{\ell} C_\ell (x_1,\cdots,x_n),
\end{split}
\end{equation}
where all the $C_\ell$ are finite because of the regulator $\Lambda$. Because of Dyson's argument \cite{dyson1952}, the series is going to be divergent and only asymptotic. Yet, and although it goes beyond the scope of the present article, it is reasonable to hope that it is Borel summable (at least for some choices of regulators) and thus provides an indirect definition of the probability measure $\upd \mu_{+\infty}$ via the expansion of its moment generating functional.

Another option to compute the properties of the beable field, inherently non-perturbative, is through a standard Monte Carlo method based on the state picture of \ref{sec:operator}. One simply evolves the state with equation \eqref{eq:psidifferential} and reweights the field probability measure with \eqref{eq:cooking} a posteriori. The issue is that \eqref{eq:psidifferential} seems extremely difficult to solve because of the rather abstract functional derivative. However, following \cite{tilloy2017}, one can introduce an auxiliary noise field $\eta$ to further decouple the non-local terms in \eqref{eq:psiintegrated} and write $\ket{\psi_\xi(t)} = \mathds{E}_\eta\big[ \ket{\psi_{\xi,\eta}(t)}\big]$ where $\ket{\psi_{\xi,\eta}(t)}$ now obeys a simple stochastic differential equation without functional derivative and hence can be computed straightforwardly. Incidentally, this provides a way to compute fermionic reduced evolutions numerically in QFT, paralleling the method of Stockburger and Grabert used for baths of harmonic oscillators \cite{stockburger2002,stockburger2004}.

\section{A possible physical theory of the world}\label{sec:world}

It is now important to show how a reasonable picture of the world emerges from the latter theory. More precisely, the objective is to show, along the lines of \ref{sec:collapseamplification}, why taking a statistical field theory of the field $\widetilde{\xi}$ gives a rewriting of QFT in realistic terms that can in principle solve the measurement problem. 

\subsection{Collapse}\label{sec:proofcollapse}
As we already have a clear local beable field $\widetilde{\xi}$, all that remains to be shown is that it undergoes a progressive collapse that is suitably amplified for macroscopic bodies. Such an analysis is easier to carry at the level of states, \ie with the formalism of \ref{sec:operator} and this is thus the picture we will subsequently use. 

We first study the evolution of a single particle to understand the collapse dynamics. Because the model is technically more difficult to handle than the CSL model of \ref{sec:collapseamplification}, we inevitably have to make restrictive assumptions that make the following derivation a hint, rather than a rigorous proof, that collapse does occur in realistic situations. In the non-relativistic limit for the fermions (hence without particle creation or annihilation) and neglecting spin, we can simplify the fermion coupling operator $\hat{\jmath}(x) \rightarrow g\, e^{i \hat{H}_0 x^0}\ket{\xb}\bra{\xb}e^{-i\hat{H}_0 x^0}:=g \,\densi (x)$ where $\hat{H}_0$ is the Schr\"odinger Hamiltonian and $\densi (x)$ has the interpretation of a density (here not mass proportional). Focusing on the pure collapse evolution and neglecting the system Hamiltonian $\hat{H}_0$ in \eqref{eq:influencefunctional} then gives a master equation very similar to that of the CSL model \eqref{eq:markovianunraveling}:
\begin{equation}\label{eq:simplemaster}
\begin{split}
\frac{\upd}{\upd t} \hat{\rho}_t \simeq - g^2 \! \int_{\mathds{R}^3}\!\!\!\!\upd^3\xb\! \int_{\mathds{R}^3}\!\!\!\!\upd^3\yb \, F(\xb,\yb,t) 
\left[\densi(\xb)[\densi(\yb),\hat{\rho}_t]\right],
\end{split}
\end{equation}
with $F(\xb,\yb,t)=\int_{0}^t\upd \tau \, D \big((t,\xb),(\tau,\yb)\big)$. Equation \eqref{eq:simplemaster} is not as nicely behaved as it might seem. Despite the appearances, it is not of the Linblad form as $F$ is not a positive semi-definite kernel. Additionally, in the absence of an explicit regularization, it is divergent because of poles at coincident points. Having the standard QFT picture in mind, this should not surprise us, equation \eqref{eq:simplemaster} contains self-energy contributions. These remarks set aside, this form of the master equation already suggests that if a localization occurs, it will be in position.

\begin{widetext}
To show that the localization does occur, one needs to study directly the stochastic evolution for states \eqref{eq:psidifferential}. With the previous approximations in mind, the latter now simply reads:
\begin{equation}\label{eq:psisimple}
\frac{\upd}{\upd t} \ket{\psi_\xi(t)} = - i g \left\{\int_{\mathds{R}^3}\!\! \upd^3\xb \; \densi(\xb)\left[\xi(t,\xb) -i g \int_{0}^t \!\! \upd^4 y \, [D-S]\left((t,\mathbf{x}),(y^0,\mathbf{y})\right)\densi(\yb)\right] \right\} \ket{\psi_\xi(t)}.
\end{equation}
Because $\densi(x)\densi(y) = \delta(x-y)\densi(x)$, we see that the evolution is diagonal in position. From now on, we take $S=0$ for simplicity but keep in mind that we will thus only get lower bounds for the collapse rate. We write $\psi_t(\xb)$ (resp. $\widetilde{\psi}_t(\xb)$) the linear (resp. normalized) wave function in position now omitting the index $\xi$. The linear wave functions can be written explicitly as a function of the history of the field $\xi$:
\begin{equation}\label{eq:psicomponents}
\psi_t(\xb) = \exp\left[-ig \int_0^t\!\!\upd \tau \, \xi(\tau,\xb) -g^2\! \int_0^t\!\!\int_0^t\!\!\upd \tau\upd s\, \theta(\tau-s)\, D\big((\tau,\xb),(s,\xb)\big)  \right]\psi_0(\xb).
\end{equation}
We now need a way to measure the fact that the state localizes on average. A reasonable metric of the collapse is given by $\Delta_t (x,y) = \sqrt{\psi^*_t(\xb)\psi_t(\xb)\psi^*_t(\yb)\psi_t(\yb)}/\langle \psi_\xi(t)|\psi_\xi(t)\rangle$. It multiplies the state density at two different points and should decrease as a function of time if the evolution indeed tends to collapse the state. Using \eqref{eq:psicomponents}, we have simply: 
\begin{equation}
\begin{split}
\Delta_t(\xb,\yb)\cdot\langle \psi_\xi(t)|\psi_\xi(t)\rangle=&\exp\left[-\frac{ig}{2} \int_0^t\!\!\upd \tau \, \xi(\tau,\xb)+\xi(\tau,\xb)-\xi^*(\tau,\xb)-\xi^*(\tau,\yb) \right]\times\\
&\exp\left[-\frac{g^2}{2} \int_0^t\!\!\int_0^t\!\!\upd \tau\upd s\, D\big((\tau,\xb),(s,\xb)\big) +D\big((\tau,\yb),(s,\yb)\big) \right]\Delta_0(\xb,\yb).
\end{split}
\end{equation}
We have chosen this measure because it makes expectation values easy to compute as the normalization factors cancel each other with the Girsanov transformed probability measure:
\begin{equation}
\mathds{E}_t[\Delta_t(\xb,\yb)] = \mathds{E}_0\left[\Delta_t(\xb,\yb)\cdot \langle \psi_\xi(t)|\psi_\xi(t)\rangle\right].
\end{equation}
We now just have to carry the Gaussian integration to get:
\begin{equation}
\mathds{E}_t[\Delta_t(\xb,\yb)] = \exp \left[\frac{g^2}{2}\!\!\int_0^t\!\!\int_0^t\!\!\!\upd \tau\upd s \, D\big((\tau,\xb),(s,\yb)\big) -\frac{D\big((\tau,\xb),(s,\xb)\big)+D\big((\tau,\yb),(s,\yb)\big)}{2}\right]\Delta_0(\xb,\yb).
\end{equation}
Writing $\Omega_t(\xb,\yb)$ the exponent, we see that the ability of the evolution to collapse a state is given by $\lim_{t\rightarrow+\infty} \Omega_t(\xb,\yb)$. The corresponding integrals are evaluated in appendix \ref{appendix:integrals} and we just discuss the results here. For a vacuum correlator $D$ and a cut-off scale $\Lambda$ we have:
\begin{equation}
\Omega_{+\infty}(\xb,\yb)=\frac{g^2}{(2\pi)^2}\left(K_0(m_b|\xb-\yb|)-K_0(\Lambda |\xb-\yb|) -\log \left[\frac{\Lambda}{m_b}\right] \right)\underset{|\xb-\yb|\rightarrow +\infty}{\longrightarrow} -\frac{g^2}{(2\pi)^2}\log \left[\frac{\Lambda}{m_b}\right],
\end{equation}
where $K_\alpha(x)$ is a Bessel function of the second kind. 
\end{widetext}

The function $\Omega_{+\infty}$ increases almost quadratically for distances smaller than $\Lambda^{-1}$, logarithmically for distances of the order of $m_b^{-1}$ and finally reaches a plateau for larger distances (see appendix \ref{appendix:integrals}). We see that if the cut-off scale is sent to infinity, all the superpositions are sharply collapsed in space regardless of their relative separation. However, for a finite cut-off, states in a spatial superposition are only partially collapsed, even for infinite times. This shows that the collapse effect is transient and dampens as times passes, at least as long as the system Hamiltonian, which does not commute with the collapse evolution, is neglected. In a more realistic situation, with many interacting particles of finite mass, one naturally expects that the collapse effect will be continuously triggered (a point we further discuss below). 

\subsection{Amplification}\label{sec:amplification}

Now that we have done a first assessment of the collapse effect, we can consider amplification. The analysis is essentially the same as that for the CSL model (see \ref{sec:collapseamplification}) and its non-Markovian extensions \cite{adler2007,adler2008}. We first need a reasonable yet tractable model of a non-relativistic many-body situation. For that matter, we consider a $N$ particle system with the same approximations as before (non-relativistic, no spin) and further assume that the effect of the Fermi statistics can be neglected so that the particles are effectively distinguishable. In this limit, we can approximate the fermionic coupling operator $\hat{\jmath}\simeq g\, \densi_{\rm tot}(\xb)=g\sum_{k=1}^N\densi_k(\xb)$ where $\densi_k$ is the standard density acting on the $k$-th particle. As before we consider the evolution of a cat state where the two peaks are initially separated by a distance much larger than the bosonic Compton wavelength $m_b^{-1}$. We also assume that the states of the particles in each peak are separated by a distance larger \footnote{This time, we cannot assume that the width of a given peak is smaller than all the length scales as we did for the CSL model for simplicity. Indeed, in that case this would mean considering distances smaller that the regularization scale which is physically unreasonable.} than $m_b^{-1}$, yet much smaller than the inter-peak distance. If we look at the corresponding density matrix this means that $\densi_k(\xb)\hat{\rho_t}\densi_l(\yb)\simeq 0$ for $|\xb-\yb|< m_b^{-1}$. The master equation \eqref{eq:simplemaster} now gives:
\begin{equation}
\begin{split}
\frac{\upd}{\upd t} \hat{\rho}_t \simeq - g^2 \sum_{k=1}^N \! \int_{\mathds{R}^3}\!\!\!\!\upd^3\xb\! \int_{\mathds{R}^3}\!\!\!\!\upd^3\yb \, F(\xb,\yb,t) 
\left[\densi_k(\xb)[\densi_k(\yb),\hat{\rho}_t]\right].
\end{split}
\end{equation}
This shows that the $N$ particles localize independently of each other thus amplifying the collapse rate of the superposition by a factor $N$. For a more in depth analysis in the close context of non-Markovian collapse models, the reader may consult the recent study of Gasbarri \textit{et al.} \cite{gasbarri2017}. 

As both the derivation of the collapse and amplification effects use rather unrealistic assumptions, ignoring effects that are likely crucial, we may provide a reassuring general argument. The stochastic unraveling we have described in this manuscript effectively hides a stochastic collapse evolution wherever there is decoherence in the usual picture. As a result, a good rule of thumb to estimate the collapse timescales is to look at the decoherence predicted by orthodox QFT. In the case of the Yukawa theory, the decoherence of fermionic degrees of freedom occurs through the emission of scalar bosons. This is why the collapse effect is only transient for a single particle. However, a macroscopic body will typically constantly emit a lot of thermal (Yukawa) radiation and thus immediately decohere, \ie collapse in our framework. Hence, even though the previous derivations have a limited scope, it is clear at a heuristic level that macroscopic objects are well localized within our theory.
 
To summarize, we have provided mathematical and heuristic arguments suggesting that our rewriting of QFT indeed solves the measurement problem through the standard collapse and amplification mechanisms. On the technical front, we have taken a non-relativistic limit for the fermionic sector but have kept the full relativistic effects of the noise with Lorentz invariant propagators (covariantly regularized). Also, where the standard analysis of collapse and amplification in a non-Markovian context \cite{adler2007,adler2008,gasbarri2017} is done perturbatively, we have studied the pure localization mechanism non-perturbatively in $g$. The purpose was incidentally to illustrate the possibility to do computations at the stochastic wave function level even with relativistic propagators.

\subsection{Empirical content and metaphysics}\label{sec:metaphysics}

We now discuss the subtle link between the empirical content of our theory (what it says about observations) and its ontological content or metaphysics (what constitutes the observations themselves). From our two derivations \ref{sec:operator} and \ref{sec:functional}, it is clear that the redefinition of QFT we have proposed is empirically equivalent to orthodox QFT. This remains true as long as the latter is understood as a tool to compute the statistics of observations. This means that, in practice, one may (and perhaps should) reuse the tools of ``standard'' QFT to compute, say, scattering cross-sections. Once the operational level has been derived from the theory, as we suggested in the previous section, there is virtually no holds barred.

On the other hand, our theory offers a rather clear and simple metaphysical picture. The only object that exists in space-time (and not in Hilbert space), is the random field $\widetilde{\xi}$. At a fundamental level, it is what the theory is ultimately about, it is what one would reasonably call ``matter''. The quantum formalism, with its Hilbert space and states, becomes simply a way to tilt the measure of this matter field away from Gaussianity. The dynamics of matter is Lorentz invariant at the deepest level in the sense that its probability measure is Lorentz invariant up to the tilt from the initial conditions (like in any other classical theory). As we have seen in the previous sections \ref{sec:proofcollapse} and \ref{sec:amplification}, the collapse and amplification mechanism tend to localize quantum states which in turn localize the field $\widetilde{\xi}$ (through the Girsanov transform of the measure). That is, macroscopic objects correspond to well localized regions of space where the time-averaged value of the fluctuating ``matter'' field $\widetilde{\xi}$ is significant.

Finally, there is no objection to apply our reformulation of QFT to the universe as a whole, without outside observers. In that case, our approach provides a probability measure from which a single ``typical'' realization is actually picked (without the need for an Everettian Many-World).  Thus, we have obtained a physical theory, written in terms of a classical random field (a statistical field theory), that can be made empirically equivalent to QFT and that can solve the measurement problem. This is no less than what we had advertised.

\section{Discussion}\label{sec:discussion}

\subsection{Reconsidering QFT}\label{sec:reconsideringQFT}
In our reformulation, a QFT is a statistical field theory for a random field $\widetilde{\xi}$ with the probability measure
\begin{equation}\label{eq:newcooked}
\begin{split}
&\upd\mu_{+\infty}(\widetilde{\xi})=\int \! \mathcal{D}[\Psi]  \mathcal{D}[\Psi']\, \rho_f(\Psi_{-\infty},\Psi'_{-\infty})\,\delta(\Psi_{+\infty},\Psi'_{+\infty}) \\
&\mathcal{Q}^*[j'] \, \mathcal{Q}[j] \exp\left(i \mathscr{S}_f-i\mathscr{S}_f' -i\int j \widetilde{\xi}-j'\widetilde{\xi}^*\right) \upd \mu_{-\infty} (\widetilde{\xi}),
\end{split}
\end{equation}
where $\upd \mu_{-\infty} $ is a Gaussian probability measure defined by its correlation and relation functions
\begin{align}
\mathds{E}_{-\infty}[\xi(x)\xi^*(y)]&=D(x,y),\\
\mathds{E}_{-\infty}[\,\xi(x)\,\xi(y)\,]&=S(x,y).
\end{align}
The purpose of the previous section was to show that this reformulation is not just a rewriting trick but that the distribution of $\widetilde{\xi}$ provides a reasonable account of the world where the measurement problem is solved. In this picture, the asymptotic fermionic quantum state tilts the measure to encode some knowledge about initial conditions.

Even if one remain unmoved by the fundamental clarification that such a rewriting entails and even if one disregards the solution to the measurement problem that naturally comes with it, one may still find some interest in the previous definition. Indeed, in this reformulation, QFT is a theory of matter from which the operational formalism only emerges. This makes the constraints on the measure $\upd \mu_{+\infty}$ of \eqref{eq:newcooked} much weaker than one would expect if the operational framework had been taken as primitive (as in standard QFT). As an illustration, we may mention the possibility to have a Lorentz invariant theory that is fundamental yet regularized in momentum. In standard QFT, regularized theories are useful for computations but they cannot be considered fundamental. A fundamental regularized orthodox QFT would require the canonical quantization of a Lagrangian with a higher degree of derivatives, but it is known that such theories cannot be defined because of the so called Ostrogradsky instability  \cite{woodward2015,motohashi2015}. Sadly, regularized orthodox QFTs are not QFTs. This is partially what motivates the study of lattice theories (where one loses Lorentz invariance but keeps a well defined theory) and String theory (that provides a natural UV regularization). We have no such issue in our case, the measure \eqref{eq:newcooked} can stay formally well defined no matter what the regularization is. In this representation, regularized QFTs can be fundamental theories. This ability to take regulators as fundamental is important because it makes the perturbative expansion encoded in equation \eqref{eq:newcooked} finite term by term and, although asymptotic, possibly Borel summable. As a result, one can entertain a very moderate yet legitimate hope that the theory may be put on mathematically rigorous grounds in the future.

Of course, if the momentum cut-off is taken to be at extremely high energy, one can still use the standard tools of renormalization at the operational level, if only to connect the bare parameters of the Lagrangian with the mass and coupling constants that are measured in experiments. In this respect, our formulation is in line with the ``modern'' view of renormalization that renormalized QFTs are effective theories of ``something''. One of the novelties of our approach is that the ``something'' can simply be a (\emph{possibly} well defined) regularized QFT.

Having a probability measure for a well defined classical field might also ease the unification with gravity. Gravitational effects may indeed simply be added in the theory by appropriately tilting the field probability measure. Alternatively, one may wish to make the field $\widetilde{\xi}$ (or more generally a function or generalization of it) gravitate in a fundamentally semi-classical theory of gravity (in the spirit of \cite{kafri2014,kafri2015,tilloy2016}). More generally, a theory of matter is much easier to marginally modify than an operational formalism. A reweighted probability measure is still always a well defined probability measure whereas a modification of a set of rules, even small, may very well make the whole edifice logically inconsistent. As a result, although operational formulations may ultimately be what one wants to make predictions, ``classical'' (or old school) theories of ``stuff'' are more suited to the construction of generalizations.

Finally, we should discuss a last methodological aspect. In the course of our derivation, we have started from an interacting QFT and proposed a new rewriting as a statistical field theory. It would be interesting to see to what extent it is possible to go the other way around and derive our formulation from first principles. Starting from a Lorentz invariant statistical field theory, what kind of assumptions would force upon us a probability measure of the form of \eqref{eq:newcooked}? Is the latter, in some reasonable sense, the simplest one can come up with? Why is biasing a Gaussian measure this way ``natural''? These look like interesting questions for further research.

\subsection{Reconsidering the dynamical reduction program}
The present article calls for an important reconsideration of the dynamical reduction program. As a matter of fact, what we have constructed can be seen as a relativistic collapse model that can be embedded, at the empirical level, in an interacting orthodox QFT. This means that a collapse model, which is more symmetric than the ones currently tested, potentially predicts no deviation from the Standard Model. As a result, the effects currently probed in experiments (see again \cite{bassi2013} for a review) can naturally be seen as artifacts of unreasonable choices of non-relativistic models. None of the standard signatures of collapse models are a necessity. The fact that collapse is expected to be empirically distinguishable from environmental decoherence comes from restrictive assumptions on the noise spectrum or on the hypothesis that intrinsic collapse cannot be shielded from. In all cases, the empirical consequences of collapse can be made to exactly match those of environmental decoherence: it is then only the properties of that environment, more or less physical, that are tested.

This admittedly surprising conclusion is ultimately related to a profound change of point of view on the physical origin of the collapse. Since the inception of collapse models, the source of the collapse mechanism has been much discussed and the origin conjectured to be a coupling with gravity \cite{diosi1989, penrose1996,pearle1996,gasbarri2017} or dark matter \cite{adler2008}. Yet the possibility that the collapse mechanism could be implemented with degrees of freedom already successfully quantized, through a stochastic unraveling, has been overlooked. What we have shown, is that collapse models can be seen as interpretations (in the sense of underlying dynamical theories with the same empirical content) rather than as modifications (in the sense of theories with a different empirical content) of quantum theory \footnote{This, however, does not mean that our approach cannot be tested. Because it allows for a wider class of probability measures than those that give rise to QFT (especially measures possibly well behaved in the UV), our statistical field theory formulation can be made to substantially deviate from QFT. If good theoretical reasons can be found for such deviations, then attempting to test them will make sense.}. 

Our approach also sheds some light on two difficulties usually encountered with collapse models: the seemingly ad hoc choice of position basis for reduction and the need for an arbitrary cut-off scale $\sigma$ in most models. The fact that collapse models reduce superpositions in position may seem ad hoc because environmental decoherence typically already selects a basis (although, of course it does not collapse states in this basis). It seems redundant to simply posit that intrinsic collapse occurs in the same basis. In our formulation, this is transparent because the same phenomenon both creates environmental decoherence and collapse. The basis is singled out by the operator that couples the fermionic and bosonic sectors and there is no further freedom involved. The need for a distance regularization has been a puzzling feature of non-relativistic collapse models as it was unclear what typical length-scale to choose. In our approach, the natural distance on which the collapse strength has substantial variations is $m_b^{-1}$, the bosonic Compton wavelength (the other length-scale $\Lambda^{-1}$ typically only changes the overall collapse strength, see appendix \ref{appendix:integrals}). This distance may look small for typical masses of the Standard Model, especially in relation with the values already experimentally or philosophically excluded. However, one should bear in mind that the collapse mechanism can be hidden in existing environmental decoherence effects and that, as a result, the existing parameter diagram \cite{feldmann2012} of authorised and forbidden values is irrelevant (at least its experimental upper-bounds).

\subsection{Limits}
The new approach to QFT and collapse models that we have discussed possesses a number of appealing features. Yet it is important to not understate the important work that still needs to be done to make the present theory totally bulletproof. 

First, many methodological aspects could be improved. Our derivation may indeed look a bit convoluted. We first trace out the bosonic sector, by Gaussian integration, then stochastically unravel it, by ``reverse'' Gaussian integration. If would certainly be more aesthetic to be able to obtain the stochastic field picture without the need for an explicit integrated representation, especially as the results of appendix \ref{appendix:decoupling} show that the results formally extend to non-Gaussian theories. As mentioned before, obtaining the final expression without starting from an interacting QFT, \ie from first principles, would be also undoubtedly more elegant. Further, our sketchy derivation of the collapse and amplification mechanisms, relying on the traditional state picture, would be more enlightening if it could be deduced directly from the functional representation of the field probability measure.

Another limitation is related to the fact that a large part of our analysis has been carried out in the restricted context of a Yukawa theory. The construction of the model in section \ref{sec:construction} is still valid for QFTs like QED up to cosmetic modifications and additional indices. However, the analysis of collapse \ref{sec:proofcollapse} and amplification \ref{sec:amplification} relies on the specific form of the fermion-boson coupling Hamiltonian. In the case of QED, in addition to the technical difficulties introduced by IR divergences, one faces the subtlety that the decoherence occurs via Bremsstrahlung \cite{diosi1990,breuer2000}. As a result, a single particle at rest would not collapse in position (even transiently). It remains reasonable to expect that the collapse would still happen in position for the center of mass of a many-body bound state but it is a fact that, at the very least, would require a more thorough analysis. Such a work is still necessary \footnote{There is of course one option that does not require any new technical analysis. It is if the reduction effect is attributed to the only existing bosonic scalar field of the Standard Model, \ie to the Brout-Englert-Higgs boson.} to compute numerical estimates of the typical timescales involved in the localization, hence in the materialization, of macroscopic bodies in Schr\"odinger cat states. An important related shortcoming of our analysis is the lack of discussion of perturbative renormalization further preventing the easy comparison of the parameters of the theory with measured values. This latter aspect would certainly deserve further analysis.

Finally, it would be interesting to have an alternative formulation of our approach in terms of ``flashes'', that is a formulation in which $\widetilde{\xi}$ is a point process instead of a field. The field $\widetilde{\xi}$ we consider indeed has very singular fluctuations and is a distribution valued object. This is neither a fundamental nor a mathematical problem, but it might yield an aesthetical discomfort. The situation cannot be improved because, it is impossible to have smooth continuous fields with Lorentz invariant probability measures. This is to be contrasted with point processes, like the sprinkling Poisson process used in Causal set theory \cite{bombelli2009}, that can have a Lorentz invariant distribution without being singular. Although appealing, the extension to such processes is not immediately trivial because there does not exist so far a stochastic jump unraveling of generic non-Markovian open-system dynamics.

\subsection{Link with previous approaches}
The present article modifies crucial assumptions, but also draws heavily, from previous approaches. Making dynamical reduction relativistic is made particularly difficult by the need for a smearing scale $\sigma$. The latter is necessary to make the collapse rate finite but spoils naive attempts at implementing Lorentz invariant dynamics. The first partially successful attempt at making collapse models relativistic without divergences has been done by Tumulka \cite{tumulka2006} in a discrete setting. An important insight of this approach is that it is much easier to make a theory Lorentz invariant at the level of the flashes (the discrete equivalent of our field $\widetilde{\xi}$) than at the level of states. Powerful in this respect, the model of Tumulka suffers from limitations: it works for non interacting particles only (in a seemingly crucial way) and uses a formalism rather far removed from what is usually used in orthodox QFT making extensions difficult. It is interesting to note that in our approach, it is the addition of interactions that is instrumental in making the collapse mechanism possible. 

The second important attack on relativistic collapse has been carried out by Bedingham \cite{bedingham2011} in the continuous case. The model of Bedingham again gives the important insight that the object that is the most naturally Lorentz invariant in collapse models is the field we have called $\widetilde{\xi}$. However, although the model is not divergent, it relies on a non-linear covariant smearing that make the statistical interpretation of the state vector subtle. Further, the model is non-Markovian in a way that makes it difficult to compute predictions, even perturbatively. Finally, the latest hint that collapse models are best understood as giving probability measures for flashes or stochastic fields has been given by Bedingham and Maroney \cite{bedingham2015,bedingham2015dice,bedingham2016} who have shown that they allow for a time-symmetric rewriting of dynamical reduction reminiscent of our equation \eqref{eq:generalized}.

Another important idea of this article is that tracing out some degrees of freedom of a QFT and then stochastically unraveling them can provide a source for dynamical reduction. Such a possibility was hinted at in a sketchy albeit prescient way by Di\'osi in the context of QED \cite{diosi2006}. The proposal was unfortunately considered ultimately unsuccessful because the theory could not be made Lorentz invariant at the level of quantum states \footnote{Although Di\'osi mentioned that other objects might be used, he did not make the proposal more concrete.}.  A further subtlety, which we mentioned previously, is that the collapse does not occur in position for a stochastic unraveling of QED, at least for single particle dynamics. 

Our construction can also be thought of as an instantiation of several ideas in foundations. The field probability measure of equation \eqref{eq:fieldcookedmeasure} can be seen as a generalized version of the signal probability measure for relativistic continuous measurement theory introduced in the seminal work of Di\'osi \cite{diosi1990relat} and as an example of the concept of a ``real'' path integral introduced by Kent \cite{kent2013}.  The idea that there exists a strong similarity between classical complex fields and quantum fields, instrumental in our unraveling, has been explored algebraically by Morgan \cite{morgan2009}. Finally, up to the small subtlety with the choice of the field measure discussed in \ref{sec:fieldmeasure}, our model takes the form of a collapse model with non-white noise, similar to that of Adler and Bassi \cite{adler2007,adler2008}. In the end, our model can be seen as a materialization of these different insights with a clarification of the resulting phenomenology (which can be made to be that of QFT).

\section{Conclusion}

We have presented a reformulation of QFT as a Lorentz invariant statistical field theory. This rewriting has been interpreted as giving an instantiation of a relativistic collapse model. This has insured that it was not just a formal tool but a plausible realistic underpinning of QFT. More precisely, the existence of localized macroscopic objects has been propounded, through the collapse and amplification mechanisms, from the statistics of the field $\widetilde{\xi}$ the theory was ultimately about. In addition to this foundational clarification, our approach has relaxed a constraint of standard QFTs by allowing for fundamentally regularized theories. This suggests that the first class (measurement) and second class (divergences) of difficulties coined by Dirac \cite{dirac1963,bell1990} might be profitably studied together. Further, we have called for a reconsideration of the dynamical reduction program by showing that collapse can easily and naturally be hidden in existing quantized fields. Collapse models do not necessarily give rise to empirical deviations from the Standard Model. Most of our analysis has been done on the example of a Yukawa theory of Dirac fermions and scalar bosons. The formal part of our reformulation would extend to other QFTs but the analysis of the collapse and amplification mechanisms, which would insure the localization of macroscopic objects, still needs to be discussed in more general settings. Finally, we have worked our way starting from what was known, that is from standard QFT, to construct our new definition of QFT as a statistical field theory. An important task for future work, now that we know that QFTs can be obtained this way, is to start directly from a Lorentz invariant statistical field theory and see what principles need to be added to get a Universe that looks like ours.

\begin{acknowledgments} 
I thank Lajos Di\'osi for igniting in me the long term reflexion that gave rise to the theory presented here. This manuscript has greatly benefited from sharp comments by Dirk-Andr\'e Deckert, Maaneli Derakhshani, Detlef D\"urr and Howard Wiseman as well as from an helpful correspondence with Gal Ben-Porath and Peter Morgan. 
\end{acknowledgments}

\appendix

\section{Computing the influence functional}\label{appendix:operatorinfluence}
Although the corresponding computation can be found in several places in the literature, we feel it is helpful to rederive the expression of the influence functional to make the present article self-contained.
The following derivation relies on Wick's theorem and is proposed \eg in Di\'osi \cite{diosi1990}, Breuer and Petruccione \cite{breuer2000} or more recently in Di\'osi and Ferialdi \cite{diosi2014}. We start from the expression of $\hat{\rho}_f(t)$ given in equation \eqref{eq:defrho}:
\begin{equation}\label{eq:defrhobis}
\hat{\rho}_f(t) = \tr_b \left\{\!\mathcal{T}\!\exp\left[ -i\!\int_{t_i}^{t} \!\!\upd^4x \, \mathscr{H}^L_{\rm int} (x)-\mathscr{H}^R_{\rm int} (x)\right] \!\cdot \!\hat{\rho} (t_i)\right\}. 
\end{equation}
We consider for simplicity that $\hat{\rho}_b$ is the vacuum state and write $\chi=\int_{t_i}^{t} \!\!\upd^4x \, \mathscr{H}^L_{\rm int} (x)-\mathscr{H}^R_{\rm int} (x)$.
Expanding the exponential and using Wick's theorem on each monomial gives:
\begin{align}
\hat{\rho}_f(t)&=\mathcal{T}_f \sum_{n=0}^{+\infty}\frac{1}{n! 2^n}  \left[- \tr_b(\mathcal{T}_b\, \chi^2 \hat{\rho}_b(t_i))\right]^n\,\cdot\hat{\rho}_f(t_i)\\
&=\mathcal{T}_f \exp\left\{-\frac{1}{2} \tr_b(\mathcal{T}_b\, \chi^2 \hat{\rho}_b(t_i)\right\}\cdot \hat{\rho}_f(t_i),
\end{align}
where $\mathcal{T}_f$ (resp. $\mathcal{T}_b$) indicates the time ordering operator for the fermionic (resp. bosonic) operators. The latter equality is easily generalized to thermal state using Wick's theorem at finite temperature \cite{evans1996}. We now only have to expand $\chi^2$ to get:
\begin{equation}
\begin{split}
\tr_b(\mathcal{T}_b\, \chi^2 \hat{\rho}_b) =&-2 \int_{t_i}^{t} \!\! \int_{t_i}^{t} \!\!  \upd^4x\, \upd^4y \; D(x,y)\, j^L(x) j^R(y) \\
&- \frac{1}{2} \theta(x^0-y^0) \, D(x,y) \, j^L(x)j^L(y) \\
&- \frac{1}{2}\theta (y^0-x^0) \, D(x,y) \, j^R(x)j^R(y),
\end{split}
\end{equation}
with $D(x,y)=\tr_b\big[\hat{\phi}(x)\, \hat{\phi}(y)\, \hat{\rho}_b (t_i)\big]$,
which leads to equation \eqref{eq:influencefunctional} of the main text. The representation of the influence functional in functional integral picture \eqref{eq:functionalphase} is then simply obtained via the standard dictionary between the operator and functional representations of quantum field theory. Notice that it may also be computed directly by Gaussian integration along the lines shown by Feynman and Vernon \cite{feynman1963}, or in a form similar to ours, in Hu and Johnson  \cite{johnson2000,johnson2002}.

\section{Complex classical noises and quantum fields}\label{appendix:gaussianfields}
The core technical result we use in this article is that the propagator of a real scalar quantum field can be written as the covariance of a (classical) complex scalar Gaussian field. Let us start by recalling the definition and basic properties of the latter object, following \cite{picinbono1996}. A Gaussian complex random variable of zero mean in finite dimension $n$ is a random vector $Z=X+iY$ of probability distribution $\mathcal{P}(z)$:
\begin{equation}\label{eq:probgaussian}
    \mathcal{P}(z) =\frac{\exp\left\{-\frac{1}{2}(z^*, z) \left(\begin{array}{cc}
        \Gamma & C  \\
        C^* &\Gamma^* 
    \end{array}\right)^{-1} \left(\begin{array}{c}
          z \\
         z^* 
    \end{array}\right) \right\}}{\pi^n \sqrt{\det (\Gamma \Gamma^* - \Gamma C^* \Gamma^{-1} C)}} 
\end{equation}
where the covariance matrix $\Gamma$ is positive semi-definite and the relation matrix $C$ is symmetric with the constraint that $P=\Gamma^* - C^* \Gamma^{-1} C$ is also positive semi-definite. These two matrices $\Gamma$ and $C$ fully characterize the process (as it is clear from eq. \eqref{eq:probgaussian}) and correspond to its two point correlation functions:
\begin{align}
    \mathds{E}[Z_i Z_j^*]&= \Gamma_{ij},\\
    \mathds{E}[Z_i Z_j]&= C_{ij}.
\end{align}
This discrete construction can formally be generalized to the continuum, \ie for fields, by functional integration (as is typically done in the physics literature). More rigorously, one may use the Bochner-Minlos theorem which guarantees that given two legitimate covariance and relation kernels, there exists a probability measure for a Gaussian process that has the latter as two point correlation functions. So all in all, if we give ourselves a positive semi-definite kernel $\Gamma(x,y)$, there exists a Gaussian (classical) $\xi$ such that $\mathds{E}[\xi(x)\,\xi^*(y)]=\Gamma(x,y)$ with some freedom remaining in the relation kernel $\mathds{E}[\xi(x)\, \xi(y)]=C(x,y)$.

Now on the quantum side, we consider $D(x,y) := \tr\left[\hat{\phi}(x)\, \hat{\phi}(y) \; \rho_0\right]$ the correlation function (not time-ordered) of a real scalar quantum field in the Heisenberg picture. The only requirement to find a probabilistic interpretation of this kernel is that it is positive semi-definite. This latter property can be seen surprisingly easily without knowing the specific expression of the propagator and is thus a general structural result. Indeed, consider a complex valued function $f$ with compact support on $\mathds{R}^4$, on has:
\begin{equation*}
\begin{split}
 (f|D|f) :=&\int\upd^4x\, \upd^4 y \; f(x) D(x,y) f^*(y) \\
 =&\tr \bigg[\underset{A^\dagger}{\underbrace{\int \upd^4 x\, f^*(x) \hat{\phi}(x)}} \, \underset{A}{\underbrace{\int \upd^4 y \, f(y) \hat{\phi}(y)}}\;\rho_0\bigg]\\
    =&\tr [ \hat{A}^\dagger \hat{A} \, \rho_0 ] \; \geq 0
\end{split}
\end{equation*}
so $D$ indeed defines a positive semi-definite kernel. 

Now, fixing simply $\Gamma=D$ we have the simple announced relation between quantum and classical complex fields:
\begin{equation}
\tr\left[\hat{\phi}(x)\, \hat{\phi}(y) \; \rho_0\right] = \mathds{E}[\xi(x)\,\xi^*(y)].
\end{equation}
It is important to note again that this does not entirely fix the stochastic process $\xi$ which still has an unspecified relation kernel $C$.
Finally, although we have done the construction with scalar fields, the result clearly extends to more complicated representations of the Poincar\'e group.

The last simple result about Gaussian fields that we use in the main text is the form of the generalized characteristic function $\varphi_\xi(a,b) = \mathds{E}\left[\exp\left(-i\int \xi a - b\xi^*\right)\right] $. Using the discrete Gaussian integration formula, one gets:
\begin{equation}
\begin{split}
\varphi_\xi(a,b) = &\exp\bigg[\iint \upd x\,\upd y \;  D(x,y) a(x)b(y)\\
&- \frac{S(x,y) a(x)a(y) + S^*(x,y)b(x)b(y)}{2}  \bigg].
\end{split}
\end{equation}
This expression can be further generalized to operator valued test fields providing a time ordering operator is added:
\begin{equation}
\begin{split}
\varphi_\xi\left(\hat{a},\hat{b}\right) =& \mathcal{T} \exp\bigg[\iint \upd x\,\upd y \;  D(x,y) \hat{a}(x)\hat{b}(y) \\
&-\frac{S(x,y) \hat{a}(x)\hat{a}(y) + S^*(x,y)\hat{b}(x)\hat{b}(y)}{2}  \bigg],
\end{split}
\end{equation}
an equality which can be proved algebraically as in \ref{appendix:operatorinfluence} using Wick's theorem (sometimes called Isserlis' theorem in this probabilistic context). This expression is then used to prove that the Ansatz \eqref{eq:psiintegrated} of the main text indeed works.

\section{Expressing the beable field \texorpdfstring{$\tilde{\xi}$}{} as a function of the Gaussian free field \texorpdfstring{$\xi$}{}}\label{appendix:girsanov}
Our objective is to show how to express $\widetilde{\xi}(t,\xb)=\xi^{[t_f]}(t,\xb)$ as a function of $\xi(t,\xb)$ when $S=0$ starting from:
\begin{equation}\label{eq:changevariablebis}
\left|\text{det} \left[\frac{\delta \xi^{[t]}}{\delta \xi}\right]\right|\, \upd \mu_t(\xi^{[t]}) = \upd \mu_{t_i} (\xi).
\end{equation}
For that matter, we shall follow the rather non-trivial steps of \cite{gambetta2002}. From \eqref{eq:changevariablebis} we have that:
\begin{equation}
\frac{\upd}{\upd t} \left\{ \left|\text{det} \left[\frac{\delta \xi^{[t]}}{\delta \xi}\right]\right|\, \upd \mu_t(\xi^{[t]})\right\}=0.
\end{equation}
Using the chain rule for differentiation and Jacobi's formula for the derivative of the determinant, the latter equality yields:
\begin{equation}\label{eq:characteristic}
\frac{\partial}{\partial t} \upd \mu_t(\xi^{[t]}) + \int \upd^4x\frac{\delta}{\delta \xi^{[t]}(x) } \left[ \upd \mu_t(\xi^{[t]}) \frac{\upd}{\upd t} \xi^{[t]}(x)\right]=0,
\end{equation}
where the partial derivative acts on the time index of $\upd \mu_t$. We now compute $\frac{\partial}{\partial t} \upd \mu_t$ using the linear stochastic Schr\"odinger equation \eqref{eq:psidifferential}:
\begin{equation}\label{eq:dev}
\begin{split}
\frac{\partial}{\partial t} \upd \mu_t=&-i\upd \mu_{t_i}\int\!\upd^3\xb\, \bra{{\psi}_{\xi^{[t]}}(t)}\, \hat{\jmath}(t,\xb)\xi^{[t]}(t,\xb) \\
+&\int_{t_i}^t\!\!\upd^4 y D((t,\xb),(y^0,\yb)) \hat{\jmath}(t,\xb)\, \frac{\delta}{\delta \xi^{[t]}(y) } \ket{{\psi}_{\xi^{[t]}}(t)} \\
+&\;\text{c.c.}.
\end{split}
\end{equation}
We note that $\ket{{\psi}_{\xi^{[t]}}(t)}$ is analytic in $\xi^{[t]}$ which allows to have the functional derivative in \eqref{eq:dev} act on the whole expectation value instead of just on kets. Further, using the functional representation of $\upd \mu_{t_i}$ provides:
\begin{equation}
\frac{\delta }{\delta\xi^{[t]}(x)} \upd \mu_{t_i}= \int_{\mathds{R}^4}\!\! \upd^4y \, D^{-1}(x,y)\, \xi^{[t]*}(y).
\end{equation}
Using this latter expression, the fact that $\upd \mu_t =\langle{\psi}_{\xi^{[t]}}(t)|{\psi}_{\xi^{[t]}}(t)\rangle \,\upd \mu_{t_i}$, and the product rule for differentiation gives:
\begin{equation}\label{eq:afterfactorization}
\begin{split}
\frac{\partial}{\partial t} \upd \mu_t=& -i \int_{t_i}^t \!\!\upd^4 y  \frac{\delta }{\delta\xi^{[t]}(y)} \\
&\times\bigg[\int \upd^3\xb D((t,\xb),(y^0,\yb)) \langle \hat{\jmath}(t,\xb)\rangle_t\, \upd \mu_t\bigg]\\
&+\text{c.c.}
\end{split}
\end{equation}
with $\langle \hat{\jmath}(t,\xb)\rangle_t=\bra{{\widetilde{\psi}}_{\xi^{[t]}}(t)}\,\hat{\jmath}(t,\xb)\,\ket{{\widetilde{\psi}}_{\xi^{[t]}}(t)}$. Finally, identifying \eqref{eq:characteristic} and \eqref{eq:afterfactorization} yields:
\begin{equation}
 \frac{\upd}{\upd t} \xi^{[t]}(x) =i \int \upd^3 \yb D((t,\yb),(x^0,\xb)) \langle \hat{\jmath}(t,\yb)\rangle_t 
\end{equation}
and thus
\begin{equation}
\xi^{[t_f]}(x)=\xi(x) +i \int_{t_i}^{t_f} \!\! \upd^4 y \,D(x,y)\, \langle \hat{\jmath}(y)\rangle_{y^0}.
\end{equation}
The compact form of the final result would seem to suggest that a simpler derivation, possibly also working for a generic $S\neq0$, should be available.

\section{Dealing with non-Gaussianities in the bosonic sector}\label{appendix:decoupling}
The objective of this appendix is to show how one can deal with anharmonicities in the bosonic action, at least at a formal level. The idea is essentially to use successive Hubbard-Stratonovich transformations to decouple higher powers in the field until the action only shows linear interaction terms. We will illustrate the reasoning on a $\phi^4$ theory.

We assume that the bosonic part of the action now reads $\mathscr{S}^{\lambda}_b=\mathscr{S}_b - \lambda\int\upd^4x\, \phi^4(x)$. As a result, the amplitude in the functional integral representation of $\rho_f$ is now multiplied by $\mathcal{A}[\phi]\mathcal{A}^*[\phi']$ with $\mathcal{A}[\phi]=\exp (-i\lambda\int\upd^4x\, \phi^4(x))$. We decouple $\mathcal{A}$ with an auxiliary field $\zeta_1$ without kinetic term to get:
\begin{equation}\label{eq:firstdecoupling}
\mathcal{A}[\phi]\propto \int\mathcal{D}[\zeta_1] \exp\left[i\int\lambda^{1/2}\zeta_1\phi^2+ \frac{1}{2} \zeta_1^2 \right].
\end{equation}
We still need to further decouple the term quadratic $\phi$ to get a linear coupling. This requires two auxiliary fields $\zeta_2$ and $\zeta_3$ because $\zeta_1$ changes of sign:
\begin{equation}\label{eq:furtherdecoupling}
\begin{split}
\exp\left[i\int\lambda^{1/2}\zeta_1\phi^2 \right]&\propto \int \mathcal{D}[\zeta_2]\mathcal{D}[\zeta_3] \exp\left[i\!\int\frac{1}{2}\left(\zeta_2^2-\zeta_3^2\right)\right]\\
&\!\!\!\!\!\!\exp\bigg[-i\!\int \lambda^{1/4}\left(\zeta_2\sqrt{\zeta_1^+}-\zeta_3\sqrt{\zeta_1^-}\right)\phi\bigg]
\end{split}
\end{equation}
where $\zeta_1^+$ (resp. $\zeta_1^-$) denotes the positive (resp. negative) part of $\zeta_1$. We now have a rewriting of the quartic amplitude $\mathcal{A}$ as a functional integral of an exponential linear in $\phi$:
\begin{equation}
\mathcal{A}[\phi]\propto\int\mathcal{D}[\zeta_1]\mathcal{D}[\zeta_2]\mathcal{D}[\zeta_3]\exp \left[i\mathscr{S}_{\rm aux}-i \, j_{\rm aux}\phi\right]
\end{equation}
where $\mathscr{S}_{\rm aux}$ is the auxiliary action quadratic in the $\zeta$'s and $j_{\rm aux}$ is the auxiliary flux implicitly defined in equation \eqref{eq:furtherdecoupling}.
In the end, the field $\phi$ is now a Gaussian field linearly coupled to $\bar{\psi}\psi$ \emph{and also} to $\zeta_1$, $\zeta_2$ and $\zeta_3$. With this rewriting, one can now proceed with the reasoning of section \ref{sec:construction} provided we still take the ground state for $\phi$ in the free theory (\ie not the true ground state). Indeed, in that case we can integrate out the $\zeta$'s doing the previous steps \eqref{eq:firstdecoupling} and \eqref{eq:furtherdecoupling} in reverse, this time with $\phi\rightarrow \xi$. This then gives a very simple non-normalized measure $\upd \nu^{\lambda}_{+\infty}(\xi)$:
\begin{equation}\label{eq:divergent}
\begin{split}
\upd\nu^{\lambda}_{+\infty}(\xi)&=\exp\left[2\, \lambda \int   \im (\xi^4)\right] \upd\mu_{+\infty}(\xi).
\end{split}
\end{equation}
As is, the measure is not normalizable because $\im (\xi^4)$ can grow faster than the quadratic Gaussian part decreases. Consequently, equation \eqref{eq:divergent} only encodes a formal perturbation expansion in $\lambda$. To really get a probability measure, one needs to cut-off large values of $\xi$, \eg with a $-\varepsilon |\xi|^6$ in the exponential.
This difficulty set aside, the fact that the potential term has been simply transported to the field $\xi$ without further modification is universal and applies to higher powers as well, probably hinting at a deeper structure.

\section{Propagator integrals}\label{appendix:integrals}
For a scalar field of mass $m$, the propagator $D^m$ reads:
\begin{equation}
D^m(x,y)=\int_{\mathds{R}^3}\frac{\upd^3\pb}{(2\pi)^3} \frac{e^{-i \sqrt{\pb^2+m^2} (x^0-y^0)+i\pb\cdot(\xb-\yb) }}{2\sqrt{\pb^2+m^2}}.
\end{equation}
From now on, we will need an explicit regularization scheme. We choose the Pauli-Villars (PV) \cite{pauli1949} regulator, that is for us the bosonic propagator will read $D=D^{m_b}-D^\Lambda$ where typically $\Lambda \gg m_b$. We note that, although this regularization is mathematically convenient, one would in principle require a scheme that preserves the positive semi-definiteness of $D$. We will assume that the final results would be qualitatively similar with such a more elaborate regularization and proceed with the PV propagators. The first integral we need to compute is:
\begin{equation}
G^m_t(\xb,\yb)=\int_0^t\!\!\int_0^t \!\! \upd \tau\upd s\, D^m\big(\tau,\xb),(s,\yb)\big).
\end{equation}
The time integration is straightforward and gives:
\begin{equation}
G^m_t(\xb,\yb) =\int_{\mathds{R}^3}\!\!\upd^3\pb \,  \frac{ e^{-i\pb\cdot(\xb-\yb)}}{(2\pi)^3} \frac{1-\cos(t\sqrt{\pb^2+m^2})}{\left(\pb^2+m^2\right)^{3/2}}.
\end{equation}
 Taking the $t\rightarrow +\infty$ limit suppresses the oscillatory cosine and allows to integrate the angular part. This gives:
\begin{align}
G^m_{+\infty}(\xb,\yb) &= \frac{- i}{(2\pi)^2 |\xb-\yb|} \int_{\mathds{R}} \upd p \frac{p\, e^{ip|\xb-\yb|}}{(p^2+m^2)^{3/2}}\\
&=\frac{1}{(2\pi)^2 }\int_{\mathds{R}} \upd p \frac{e^{ip|\xb-\yb|}}{\sqrt{p^2+m^2}}\\
&=\frac{2}{(2\pi)^2}K_0\left(m\,|\xb-\yb|\right),
\end{align}
where $K_{\alpha}(x)$ is a modified Bessel function of the second kind. Defining $G=G^m-G^\Lambda$, we can finally proceed to the computation of the integral $\Omega_\infty$ that was needed in \ref{sec:proofcollapse}:
\begin{align}
&\Omega_{+\infty}(\xb,\yb)= \frac{g^2}{2} \Big[G_{+\infty}(\xb,\yb)\!-\!\frac{G_{+\infty}(\xb,\xb)+G_{+\infty}(\yb,\yb)}{2}\Big]\\
&=\frac{g^2}{(2\pi)^2}\bigg(\!K_0\left(m_b|\xb-\yb|\right)-K_0\left(\Lambda|\xb-\yb|\right) - \log\left[\frac{\Lambda}{m_b}\right]\!\bigg).
\end{align}
It is an increasing function of $|\xb-\yb|$ that has the following asymptotic behavior:
\begin{align}
\Omega_{+\infty}(\xb,\yb) &\underset{ |\xb-\yb|\gg m_b^{-1}\gg\Lambda^{-1}}{\sim} -\frac{g^2}{(2\pi)^2}\log\left[\frac{\Lambda}{m_b}\right],\\
\Omega_{+\infty}(\xb,\yb) &\underset{m_b^{-1}\gg |\xb-\yb|\gg \Lambda^{-1}}{\sim} - \frac{g^2}{(2\pi)^2} \log\left(|\xb-\yb| \Lambda\right),\\
\Omega_{+\infty}(\xb,\yb) &\underset{m_b^{-1}\gg \Lambda^{-1}\gg |\xb-\yb|}{\sim} \frac{g^2}{4(2\pi)^2}\Lambda^2  |\xb-\yb|^2\log |\xb-\yb|.
\end{align}

\bibliographystyle{apsrev4-1}
\bibliography{main}
\end{document}